\documentclass[runningheads]{llncs/llncs}
\usepackage[utf8]{inputenc}
\usepackage{cite}
\usepackage{hyperref}
\usepackage{wrapfig}
\usepackage{listings}
\usepackage[ruled]{algorithm2e}
\usepackage{tikz}
\usepackage{amsmath}
\usepackage{amsfonts}
\usepackage{float}
\usepackage{pgfplots}
\usepackage{subcaption}
\usepackage{lineno}
\usepackage{calc}
\usepackage{etoolbox}
\usepackage{float}
\usepackage{array}
\floatstyle{boxed}
\newfloat{boxed_table}{h}{modal_props.aux}
\floatname{boxed_table}{Table}

\pgfplotsset{compat=1.17}  
 
\newcommand{\sequential}{\mathbin{.}}
\newcommand{\boolsort}{\mathit{Bool}}
\newcommand{\ie}{\emph{i.e.}}
\newcommand{\eg}{\emph{e.g.}}
\newcommand{\natsort}{\mathit{Nat}}
\newcommand{\var}[1]{\ensuremath{\textit{#1}}}
\newcommand{\cpp}{\texttt{C++}}

\newcommand{\tab}[2]{\setlength{\hspace*}{0.4cm * #2}#1\setlength{\hspace*}{0.4cm - \widthof{#1}}}
\newcommand{\true}{\textit{true}}
\newcommand{\false}{\textit{false}}
\newcommand{\Nat}{\mathbb{N}}
\newcommand{\arity}[1]{\mathit{ar\!}_{#1}}  

\title{A thread-safe Term Library
\thanks{Supported by projects 612.001.751 (NWO, AVVA) and 00795160 (TTW, MASCOT).} 
\vspace{0.3ex}\\
\small(with a new fast mutual exclusion protocol)~~~~}
\titlerunning{A thread-safe Term Library}

\author{Jan Friso Groote\orcidID{0000-0003-2196-6587}  
\and
Maurice Laveaux\orcidID{0000-0001-8732-7580} 
\and
P.H.M. van Spaendonck\orcidID{0000-0002-9536-1524}
}

\authorrunning{J.F. Groote, M. Laveaux and P.H.M. van Spaendonck}

\institute{Department of Mathematics and Computer Science, Eindhoven University of Technology, Eindhoven, The Netherlands\\
\email{\{J.F.Groote,M.Laveaux,P.H.M.v.Spaendonck\}@tue.nl}}

\date{}

\begin{document}
\maketitle
\begin{abstract}
\noindent Terms are one of the fundamental mathematical concepts in computing. 
E.g.\ every expression characterisable by a context free grammar is a term. 
We developed a \textit{thread-safe} Term Library. The biggest challenge is to 
implement hyper-efficient multi-reader/single-writer mutual exclusion for which we designed the new 
\textit{busy-forbidden protocol}. Model checking is used to show both the correctness of the
protocol and the Term Library. Benchmarks show this Term Library has little overhead compared to 
sequential versions and outperforms them already on two processors. 
Using the new library in an existing state space generation tool, 
very substantial speed ups can be obtained. 

\keywords{Term Library \and Mutual exclusion \and Thread-safe \and Model checking.}
\end{abstract}

\section{Introduction}
A term is a common mathematical structure. Many concepts can be represented
as terms, such as programs, specifications and formulas. Many operations
in computing are term transformations, such as compilation. In computer science
a term is a far more commonly used concept than structures such as arrays, lists or matrices. 
This makes it remarkable that terms are not a standard data structure in common programming
languages such as $\cpp$ and Java.  

To our knowledge the first term library stems from the realm 
of program transformations. In \cite{DBLP:conf/coordination/BergstraK96, DEJONG200435,ip-sen_20041181, aterm,VANDENBRAND200755} an ATerm library of so called \textit{annotated terms} has been proposed, 
which contains terms with meta information. Stripping away all bells and whistles from this ATerm format, 
a very plain and elegant term data structure remains. 

Our terms are defined in the standard way. We start out with a given set of function symbols $F$ 
where each function symbol $f\in F$ has an arity $\textit{ar}_f$. 
Each constant function symbol, i.e.\ with arity $0$, is a term. Given a function 
symbol $f\in F$ with $\arity{f}>0$, and terms 
$t_1,\ldots, t_{\arity{f}}$, the expression $f(t_1,\ldots,t_{\arity{f}})$ is
also a term. These are the only two ways to construct a term. 

As an example, we provide terms where some constants represent variables. We can have function
symbols $\{0,1,x,y,+\}$ and have terms $0+1$, $x+1$ and $x+y$. The `constants' $x$ and
$y$ allow for different operations than the constants $0$ and $1$, as it is natural
to define a substitution operation for the constant $x$, whereas that would be
less natural for the constant $0$. In a similar way, terms with binders can be
represented. For instance, in the term $\lambda x.t$ the $\lambda$ is just a binary
function symbol and the first subterm is the variable $x$.

As in the ATerm library, terms are stored in a maximally shared way. 
Once created, terms remain as stable structures in memory until they are garbage collected. 
This leads to a smaller memory footprint, because equal terms are only
stored once. Comparing terms is also computationally cheap, as two terms are equal iff they occupy the
same address in memory. Also note that handing a term over to another thread is
also cheap, as only the address of the term needs to be transferred. This avoids serialising
and deserialising terms as done in~\cite{DBLP:journals/logcom/BlomLP011}.
A disadvantage is that subterms cannot be replaced. If a subterm needs to be changed,
the whole surrounding term must be reconstructed.  

With a steadily increasing number of computational cores in computers, 
it is desirable to have a parallel implementation of a term library. 
As terms have a tree-like structure, one would expect concurrent tree algorithms, 
as provided by the EXCESS project \cite{excess} or the PAM library \cite{PAM}, to be a useful solution. 
However, these tree libraries concentrate on manipulating the trees themselves, by adding and removing nodes, and
rebalancing when required. This would not allow maximal sharing of terms, which have to be static structures
in memory. 

Early attempts to create a thread-safe term library led to intriguing wait-free algorithms
\cite{DBLP:journals/dc/HesselinkG01,DBLP:journals/dc/GaoGH05,DBLP:journals/scp/GaoGH07}.
The assumption was that
thread synchronisation was the root cause of performance issues, and this is avoided when
algorithms are wait-free.
But this did not turn out to be entirely true. 
As the operations to create, inspect and destroy terms occur frequently and 
are computationally very cheap, a thread-safe implementation allows for hardly any overhead.  
Wait-free algorithms are intricate and their overhead is deadly for performance in this case. 
The same applies to the introduction of mutex variables 
surrounding construction, inspection and deletion of terms. 

Although the need and advantages of having terms that can be accessed by multiple threads
have already been stressed in the original publications, it turns out to be hard to make
a thread-safe term library that is competitive with sequential implementations. This is most likely the reason
that no thread-safe term libraries exist, except for a non-published Java implementation \cite{LankampParAterm}.

In this article we present a thread-safe term library that is competitive with 
sequential term libraries. We first observe that with some minor adaptations, \ie, essentially
introducing a Treiber stack~\cite{treiber1986systems} in a hash table, inspection and construction 
of terms can happen concurrently.
Secondly, we note that garbage collection on the one hand, 
and construction/moving/copying of terms on the other hand
must be mutually exclusive, 
and construction happens far more often than garbage collection. 

Therefore, we require a mutual exclusion algorithm with behaviour of a readers-writer lock
\cite{DBLP:books/daglib/0030596}, where construction of terms can happen simultaneously (=readers), and 
garbage collection (=writer) must be done in isolation. 
However, standard readers-writer locks are too expensive.
We designed the \textit{busy-forbidden protocol} that 
employs this asymmetric access pattern as well as the
cache structure of modern processors. Obtaining access to construct a term only requires access to
two bits, virtually always available in the local cache of the current processor. 
Besides this, we developed thread-safe term protection mechanisms, either using atomic operations
for reference counting, or by employing explicit thread-local protection sets.

Experiments show that the new Term Library scales well and for practical tasks it is already beneficial when only two processors are available. The solution with a standard
readers-writer lock and especially the Java implementation are substantially slower than our
implementation with the busy-forbidden protocol. 

The correctness of thread-safe implementations is subtle. Therefore, we use the mCRL2 model checking
toolset \cite{groote2014modeling}
to design both the busy-forbidden protocol and the Term Library, and prove their correctness properties,
before implementation. This turned out to be very effective, as we did not have to struggle with
obscure faults due to parallel behaviour in the algorithm. It is intended that the new thread-safe 
Term Library will form the heart of the new release of the mCRL2 toolset. The currently existing early prototype
already achieves speed ups of a factor 12 on 16 processors for a computationally intensive task,
namely state space generation, which is more than just promising. 

\section{The term data structure}
In \cite{aterm,VANDENBRAND200755} a term library has been proposed. A term is 
a very frequently used concept within computer science.
The original motivation for terms as a basic data structure came from 
research in software transformation \cite{DEJONG200435,ip-sen_20041181}. The model checking toolset 
mCRL2 uses terms to represent all internal concepts, such as modal formulas, 
transition systems and process specifications \cite{groote2014modeling}. 

\subsection{The external behaviour of the Term Library}\label{section:atermpp_interface}
Terms are constructed out of \textit{functions symbols}, or for short \textit{functions}, 
from some given set $F$. 
Each function $f\in F$ has a number of arguments $\arity{f}$, generally called 
the \textit{arity} of $f$. A function symbol with arity $0$ is called a \textit{constant}.

\begin{definition} 
\label{def:terms}
Let $F$ be a set of function symbols.
The set of terms $T_F$ over $F$ is inductively defined as follows:
\[ \textrm{if }f\in F, f\textrm{ has arity }
\arity{f} \textrm{ and }t_1,\ldots,t_{\arity{f}}\in T_F, \textrm{ then }f(t_1,\ldots, t_{\arity{f}})\in T_F.\]
\end{definition}

\begin{wrapfigure}{r}{0.22\textwidth}
    \centering
    \vspace{-2ex}
    \begin{tikzpicture}
        \node (f0) at (0.5,0) {$+$};
        
        \node(3) at (0,-0.8) {$3$};
        \node(f1) at (1,-0.8) {$*$};
        
        \node(4) at (0.5,-1.6) {$4$};
        \node(g) at (1.5,-1.6) {$2$};
        
        \draw (f0) -- (3);
        \draw (f0) -- (f1);
        \draw (f1) -- (4);
        \draw (f1) -- (g);
    \end{tikzpicture}
    \caption{The tree for $3+4*2$.}
    \label{fig:termtree}
    \vspace{-4ex}
\end{wrapfigure}
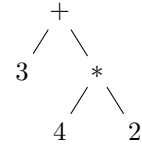

Simple numeric expressions are typical examples of terms. The function symbols are 
$0,1,2,3,+,*$ where $0, 1, 2, 3$ are constants and $+$ and $*$ have arity $2$. 
An example of a term as a tree structure is given in Figure \ref{fig:termtree}.

The term library in \cite{aterm,VANDENBRAND200755} allows to annotate terms, hence 
the name ATerm, but we do not use this feature. This original ATerm proposal also 
supported special terms representing numbers, strings, lists and even `blobs' containing
arbitrary data. We made our own implementation of a term library where besides terms as defined in
Definition \ref{def:terms},
there are also facilities for lists and 64-bit machine numbers. As these are in many respects the same as
terms constructed out of function symbols, we ignore lists and numbers in this exposition. 

From the perspective of a programmer terms are immutable 
maximally shared tree structures in memory. This means that if two (sub)terms are the same, they
are represented by the same address in memory. The term library provides essentially the following
limited set of operations on terms:

\begin{trivlist}
\setlength{\itemsep}{0cm}
\item[]\textbf{Create.}
Given a function symbol $f$ and terms $t_1,\ldots, t_{\arity{f}}$ construct a term
$f(t_1,\ldots,t_{\arity{f}})$. This operation can fail when there is not enough memory. 
\item[]\textbf{Destroy.}
Indicate that a term $t$ will not be accessed anymore by this thread. Terms that are not accessed by any thread must
ultimately be garbage collected. 
\item[]\textbf{Copy/move.}
Move or copy a term. This essentially means move or copy the address of the term. 
\item[]\textbf{Argument.}
Obtain the $i$-th subterm $t_i$ of a term $f(t_1,\ldots,t_{\arity{f}})$.
\item[]\textbf{Function.}
Obtain the function symbol $f$ of a term $f(t_1,\ldots,t_{\arity{f}})$.
\item[]\textbf{Equality.}
For terms $t$ and $u$ determine whether $t$ and $u$ are equal. Note that
due to maximal sharing this operation only requires constant time. 
\end{trivlist}

Due to the immutable nature of terms in memory it is not possible to simply replace a subterm of a term.
If a subterm must be changed, the whole surrounding term must be copied. On the other hand terms are
very suitable for parallel programming. Threads can safely traverse protected terms in memory as the treads can
be sure that these terms will not change.   

By storing terms as maximally shared trees, the only non trivial operations on terms 
are the creation of a new term and the destruction of an existing term. 
Given a function symbol $f$ and subterms $t_1,\ldots,t_{\arity{f}}$ it
must be determined whether the term $f(t_1,\ldots,t_{\arity{f}})$ already exists. This is done using
a hash table. If the term already exists, this term
is returned. If not, a new term node labelled with $f$ pointing to the subterms $t_1,\ldots,t_{\arity{f}}$ 
must be made. 

The typical usage pattern of terms is that they are visited very often obtaining arguments or 
function symbols. Creation of a term is also a very frequent operation, where in the majority of
cases a term is created that already occurs in the hash table. Only rarely a garbage collect is taking place. 

\subsection{Behavioural properties of the Term Library}
\label{behaviouralproperties}
The Term Library guarantees the following properties, checked using model checking, see Section \ref{section:busy_forbidden_verification}.
\begin{enumerate}
\item \label{item:prop_no_spontaneous_destroy}
A term and all its subterms remain in existence at exactly the same 
address, with unchanged function symbol and arguments, as long as it is not destroyed. 
\item \label{item:prop_injective}
Two stored terms $t_1$ and $t_2$ always have the same non-null address iff they are equal. 
\item \label{item:prop_can_start}
Any thread that is not busy creating or destroying a term, can always initiate the construction of a new term or the destruction of an owned term.
\item\label{item::finish_eventually}
Any thread that started creating a term or destroying a term, will eventually successfully finish this task provided there is enough memory to store one more term than
those that are in use. But it is required that other threads behave fairly, in the sense
that they will not continually
create and destroy terms or stall other threads by busy waiting. 

\end{enumerate}
Note that the properties above imply some notion of garbage collection in the
sense that if a thread makes and destroys terms, and these are not garbage collected,
at some point no new terms can be created due to a lack of memory and in that case property
\ref{item::finish_eventually} above would be violated. 

\subsection{The implementation of the thread-safe Term Library}
Terms are implemented in the Term Library by storing them in a hash table. 
Whenever a term with function symbol $f$ and arguments $t_1,\ldots,t_{\arity{f}}$ is created,
the hash table is used to find out whether $f(t_1,\ldots,t_{\arity{f}})$ already exists. 
If yes, its current address is returned. If no, a new term $f(t_1,\ldots,t_{\arity{f}})$, is inserted
in the hash table and its address is returned. 

Another possible solution would be to use a CTrie \cite{Ctrie} instead of the hash table. 
However CTries main advantage, memory conservation, over performance, makes it less 
suitable for our Term Library, which must be suitable to deal with huge numbers of term
manipulations in short time spans. 

Terms in our Term Library can be constructed and accessed in parallel. When a thread 
created a term, this term and all its subterms are immutable and stored at fixed addresses in memory, and
this means that any term can be accessed safely by all threads that have not destroyed the term.

We have two ways to implement garbage collection in the thread-safe Term Library,
namely reference counting and the use of protection sets, which ensure that non-destroyed terms remain in memory. 
Garbage collection is performed by a single thread. Note that
mark-and-sweep algorithms exist where creation and destruction can be done simultaneously
\cite{DBLP:journals/scp/GaoGH07} but these are very complex. As garbage collection is relatively 
fast, such advanced algorithms are not necessary. 

In reference counting,
each term has a reference count that is incremented by one whenever a term is created or copied,
and decremented by one if a thread drops a reference to the term.
Terms that are not in use anymore have a reference count of zero and can be garbage collected. 
This can easily be performed by visiting all terms, which are stored in traversable structures.

An alternative is to use term protection sets. Whenever a term is stored at some address,
this address is stored in a separate protection set, locally maintained by each
thread. When the address is not used anymore for a term, it is removed from the set. As
every address can only be stored once, a simple hash table suffices to implement the protection
set. Garbage collection consists of marking all terms reachable via some protection set, and
removing all others. 

In the parallel setting changing reference counts or inserting/deleting addresses in protection sets
must be sequentially consistent meaning that they cannot be rearranged in the programs. 
Changing reference counts must be atomic and can lead to cache contention 
as the reference counts are accessible by all threads. 
Operations on the protection sets are far more complex
than changing a reference count, but they are always local in a thread, and depending on the style of
programming need to be executed far less often than changing a reference count. From the benchmarks we 
derive that protection sets are preferable. 

If we only create terms, this can be done in parallel as well. We use a dedicated hash table
with a bucket list in the form of a linked list to check whether a term already exists. 
If the term does not exist, it is added using a compare and swap operation to the bucket
list of the appropriate entry of the hash table. If in the mean time another thread creates
the same term, the compare and swap fails, informing the thread that it has to 
inspect the hash table again to find out whether the term came into existence. 
This is Treiber's stack, which is sufficient since terms are not
simultaneously deleted from the bucket lists. Deletion only occurs during 
garbage collection, and during garbage collection no new terms are allowed to 
be constructed. 

Accessing terms during garbage collection and rehashing is perfectly safe. 
But it is not allowed to create or copy terms while garbage collection or rehashing is going on. 
This requires a mutual exclusion protocol where either multiple threads can create and copy terms simultaneously, which we call the \textit{shared} tasks, or a single thread can be involved in garbage collection or rehashing, which is called the \textit{exclusive} task. 

This is the same as a readers-writer lock \cite{DBLP:books/daglib/0030596} where multiple
readers or at most one writer can access a shared resource. Reading is the shared task, and 
writing is exclusive. 
As we observed that creating and copying terms is done very frequently compared to garbage collection, 
shared access must be cheap and exclusive access can be expensive. Most standard 
readers-writer locks require at least one access to a common mutex variable for shared 
access which is so costly that parallel implementations based on the readers-writer lock run on multiple processors
failed to outperform the sequential implementation. This observation is supported by the benchmarks. 
We developed a completely new protocol, called the \textit{busy-forbidden}
protocol serving our needs, which is described in the next section.

\begin{table}[h]
    \centering
~~\begin{tabular}{|l|l|}
\hline
\begin{minipage}[t]{0.62\textwidth}
\resetlinenumber
\begin{internallinenumbers}
$\texttt{create}(\textit{thread}~p,~\textit{symbol}~f,~\textit{subterms}~t_1, \ldots, t_n)$\\
\hspace*{0.4cm}$\texttt{enter\_shared}(p);$\\
\hspace*{0.4cm}$\textit{hash}:=\texttt{h}(f,t_1, \ldots, t_n);$\\
\hspace*{0.4cm}$\textit{bucket}:=\textit{buckets}[\textit{hash}]$;\\
\hspace*{0.4cm}$t := \texttt{insert}(\textit{bucket},f,t_1,\ldots,t_n)$;\\
\hspace*{0.4cm}$\texttt{protect}(p,t)$;\\
\hspace*{0.4cm}$\texttt{leave\_shared}(p);$\\
\hspace*{0.4cm}$\textbf{return}~t;$\\
\\
$\texttt{insert}(\textit{bucket}~b,~\textit{symbol}~f,~\textit{subterms}~t_1, \dots, t_n)$\\
\hspace*{0.4cm}$old\_head,node := b.top;$\\
\hspace*{0.4cm}$\textbf{do}$\\
\hspace*{0.8cm}$\textbf{if}~node.head~\textbf{represents}~f(t_1, \dots, t_n)$\\
\hspace*{1.2cm}$\textbf{return}~node.head;$\\
\hspace*{0.8cm}$\textit{node} := \textit{node.tail};$\\
\hspace*{0.4cm}$\textbf{while}~(node \neq \texttt{NULL});$\\
\hspace*{0.4cm}$t := \textbf{construct}~f(t_1, \ldots, t_n);$\\
\hspace*{0.4cm}$\textbf{if not}~\texttt{cmpswap}(b.top,old\_head, Node(t,old\_head))$\\
\hspace*{0.8cm}$\textbf{destruct}~t$;\\
\hspace*{0.8cm}$\textbf{return}~\texttt{insert}(b, f, t_1, \ldots, t_n);$\\
\hspace*{0.4cm}$\textbf{return}~t;$
\end{internallinenumbers}
\end{minipage}
     &
\begin{minipage}[t]{0.30\textwidth}
$\texttt{destroy}(\textit{thread}~p,~\textit{term}~t)$\\
\hspace*{0.4cm}$\texttt{unprotect}(p);$\\
\hspace*{0.4cm}$\textbf{possibly do}~\texttt{GC}(p)$\\
\\
$\texttt{GC}(\textit{thread}~p)$\\
\hspace*{0.4cm}$\texttt{enter\_exclusive}(p);$\\
\hspace*{0.4cm}$\textbf{forall}~t \in hash\_table$\\
\hspace*{0.8cm}$\textbf{if not}~\texttt{protected}(t)$\\
\hspace*{1.2cm}$\textbf{remove}~t$;\\
\hspace*{0.4cm}$\texttt{leave\_exclusive}(p);$\\
\end{minipage}\\
\hline
\end{tabular}
    \caption{Pseudocode description of the thread-safe Term Library.}
    \label{table:par_aterm}
\end{table}

Using the busy-forbidden protocol, a compare and swap to insert terms in bucket lists for the 
hash table, the implementation of thread-safe Term Library is pretty straightforward but delicate.
Table \ref{table:par_aterm} contains the code for creating and destroying terms. In this code \texttt{enter\_shared}, \texttt{leave\_shared}, \texttt{enter\_exclusive} and \texttt{leave\_exclusive} are part of the busy-forbidden protocol described in the next section.  
The function \texttt{h} is a hash function that takes a function symbol \textit{f}, and subterms $t_1, \ldots, t_n$, and calculates a possibly non-unique hash. The functions \texttt{protect}, \texttt{unprotect} and \texttt{protected} refer to the protection mechanisms described earlier, in which $\texttt{protected}(t)$ will return true if and only if the term $t$ is protected by some thread. 
Besides this, each bucket $b$ in the hash table contains an atomic pointer $\textit{b.top}$ that 
allows atomic loads and an atomic compare-and-swap operation \texttt{cmpswap}, which returns true if and only if successful.
The call $\texttt{GC}(p)$ stands for doing a garbage collect by thread $p$.

Using an mCRL2 model of the behaviour of the Term Library, 
the behavioural properties mentioned in Section \ref{behaviouralproperties} have been
model checked. This is described in Section \ref{section:busy_forbidden_verification}. 

\section{The busy-forbidden protocol}\label{section:busy_forbidden}
The busy-forbidden protocol is of independent interest. 
This protocol guarantees that at most one thread can be in state \textit{Exclusive} and if a thread is
in state \textit{Exclusive}, no thread is in state \textit{Shared}, and vice versa, if
there are threads in state \textit{Shared}, then there is no thread in the state \textit{Exclusive}. 
It behaves in a similar way as a readers-writer lock \cite{DBLP:books/daglib/0030596}, called a shared mutex in
$\cpp$. 

The busy-forbidden protocol is designed for the situation where 
shared access is frequent whereas exclusive access is infrequent.

\subsection{The external behaviour of the busy-forbidden protocol}
We first look at the external behaviour of this protocol. As indicated above, threads can request for shared or
exclusive access by calling one of the two functions \texttt{enter\_shared} and \texttt{enter\_exclusive}. 
The functions starting with \texttt{leave} are used to indicate that access is no longer required.

\begin{figure}[t]
\begin{center}
\newcommand{\state}[4]{
     \node[thick, circle, minimum size=0.1cm, draw] (#3) at (#1,#2) 
     {\raisebox{0cm}[1ex][0ex]{\makebox[0.8cm]{\small\textit{#4}}}};}
\newcommand{\trans}[4]{\draw[thick, ->] (#1) -- (#2) node[midway, rotate=#4]{\small\begin{tabular}{c}#3\end{tabular}};}
\newcommand{\tauloop}[2]{
    \draw[->, thick] (#1) 
        edge[in=\ifthenelse{\ifstrequal{#2}{left}}{45}{225}, out=\ifthenelse{\ifstrequal{left}{#2}}{-45}{-224}]
        node[#2]
        (#1);
}
\begin{minipage}{0.48\textwidth}
\resizebox{\textwidth}{!}{\begin{tikzpicture}
  \state{ 0.00}{0.00}{Free}{Free};
  \state{-1.90}{1.38}{EnteringShared}{EnterS}
  \state{-1.18}{3.61}{LockedOutExclusive1}{LOE1}
  \state{ 1.18}{3.61}{Shared}{Shared}
  \state{ 1.90}{1.38}{LeavingShared}{LeaveS}
  \trans{Free}{EnteringShared}{\texttt{Enter}\\\texttt{shared}\\\texttt{call}}{54}
  \trans{EnteringShared}{LockedOutExclusive1}{No threads in\\ \textit{LOS} or \textit{Exclusive}}{-18}
  \draw[->, thick] (EnteringShared)
    edge[in=225, out=-225, looseness=3.4]
    node[above, rotate = 90]
    {\begin{tabular}{c}
        \textit{improbable} \\
        At least 1 thread\\
        in \textit{LOS} or \textit{Exclusive}
    \end{tabular}}
    (EnteringShared);
  \trans{LockedOutExclusive1}{Shared}{\texttt{Enter}\\\texttt{shared}\\\texttt{return}}{90}
  \trans{Shared}{LeavingShared}{\texttt{Leave}\\\texttt{shared}\\\texttt{call}}{18}
  \trans{LeavingShared}{Free}{\texttt{Leave}\\\texttt{shared}\\\texttt{return}}{-54}
  \state{+2.17}{-1.25}{EnteringExclusive}{EnterE}
  \state{+2.17}{-3.75}{LockedOutExclusive2}{LOE2}
  \state{+1.18}{-6.63}{LockedOutAll}{LOS}
  \state{-1.18}{-6.63}{Exclusive}{\scriptsize Exclusive}
  \state{-2.17}{-3.75}{LeavingExclusive1}{\scriptsize LeaveE1}
  \state{-2.17}{-1.25}{LeavingExclusive2}{\scriptsize LeaveE2}
  \trans{Free}{EnteringExclusive}{\texttt{Enter}\\\texttt{exclusive}\\\texttt{call}}{60}  
  \trans{EnteringExclusive}{LockedOutExclusive2}{No threads in\\\textit{LOE2}, \textit{LOS}, \textit{LeaveE1}\\or \textit{Exclusive}}{0}  
  \draw[thick, ->] (LockedOutExclusive2) edge[in=45, out=-45, looseness=3.4] node[above, rotate = -90] {\textit{improbable}} (LockedOutExclusive2);
  \trans{LockedOutExclusive2}{LockedOutAll}{No threads in\\\textit{LOE1} or \textit{Shared}}{-21}  
  \trans{LockedOutAll}{Exclusive}{\texttt{Enter}\\\texttt{exclusive}\\\texttt{return}}{90}
  \trans{Exclusive}{LeavingExclusive1}{\texttt{Leave}\\\texttt{exclusive}\\\texttt{call}}{21}
  \trans{LeavingExclusive1}{LeavingExclusive2}{$\tau$~~~~}{0}
  \draw[thick, ->] (LeavingExclusive1) edge[in=225, out=-225, looseness=3.4] node[above, rotate = 90] {\textit{improbable}}(LeavingExclusive1);
  \trans{LeavingExclusive2}{Free}{\texttt{Leave}\\\texttt{exclusive}\\\texttt{return}}{-60}  
\end{tikzpicture}}
\end{minipage}
\begin{minipage}{0.46\textwidth}
\resizebox{\textwidth}{!}{
\begin{tabular}{ll}
\\
\textit{LOE1}&There are no threads\\
    &in or able to enter \textit{Exclusive}.\\
\textit{Shared}&Shared access. No concurrent\\
&access to \textit{Exclusive} possible.\\
\\
\textit{EnterS}&Entering shared.\\
\textit{LeaveS}&Leaving shared.\\
\\
\textit{Free}&The thread is outside any\\
&exclusive or shared section.\\
\\
\textit{LeaveE2}&Leaving exclusive.\\
\textit{EnterE}&Entering exclusive.\\
\\
\textit{LeaveE1}&Leaving exclusive. No threads \\
&in or able to enter \textit{Exclusive}.\\ 
\textit{LOE2}&There are no threads in\\
& or able to enter \textit{Exclusive}.\\
\\
\textit{Exclusive} & Exclusive access. There are\\
& no threads in or able to enter \\
& \textit{Exclusive} or \textit{Shared}.\\
\textit{LOS}&No threads in or able to enter\\
& \textit{Exclusive} or \textit{Shared}.\\
\end{tabular}\vspace*{1cm}}
\end{minipage}
\end{center}
\caption{The external behaviour of the busy-forbidden protocol.}
\label{fig:external-busy-forbidden}
\end{figure}
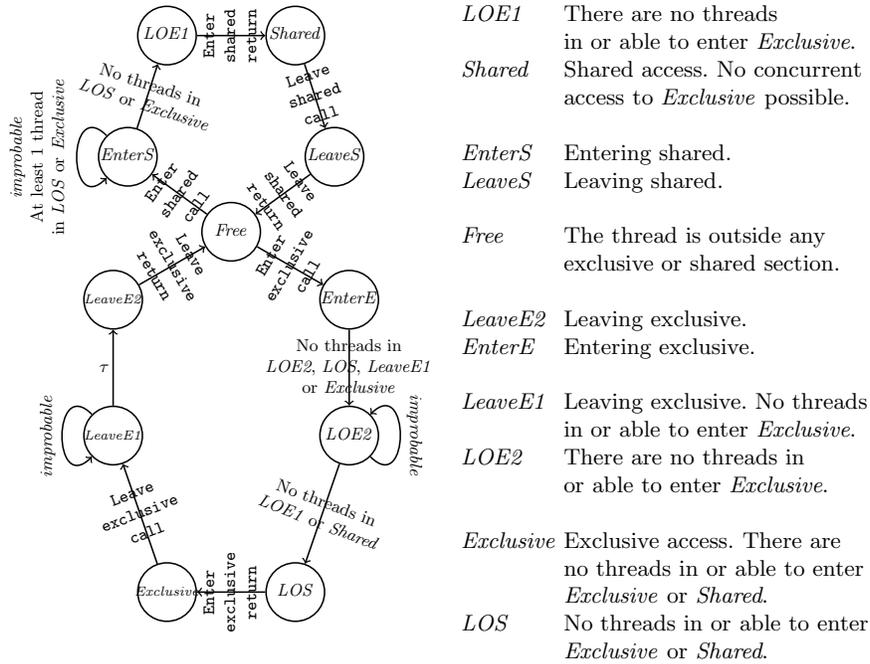

We make the external behaviour more precise by modelling it as a state automaton, actually obtained
by the specification in mCRL2 used for verification. From the perspective of a single thread, the behaviour is depicted in Figure~\ref{fig:external-busy-forbidden}.
The calls are modelled by actions \texttt{Enter}/\texttt{Leave} \texttt{shared/exclusive} \texttt{call}. 
Returning from the function is modelled by actions ending in \texttt{return}.

The centre state, marked \textit{Free}, indicates that the thread is not involved in the protocol. 
It is outside the shared and exclusive sections. Following the
arrows in a clockwise fashion, a thread obtains
access. In the state \textit{EnterS} the thread requested shared access, and it will get
it when there are no threads in the states \textit{LOS} or \textit{Exclusive}. 
From the figure it is quite easy to see that the protocol indeed satisfies the mutual exclusion 
constraints mentioned above. 

We went to great length to ensure that the behaviour of 
Figure~\ref{fig:external-busy-forbidden} for multiple threads
is divergence-preserving branching bisimilar
to the implementation below \cite{DBLP:journals/jacm/GlabbeekW96,GLT09a}. 
This equivalence is equal to branching bisimulation, but it 
does not remove $\tau$-loops, \ie, loops of internal actions.
It preserves not only safety but also liveness properties, 
and allows us to use this specification to verify the Term Library.

The loop at \textit{EnterS} occurs typically when another thread is in state \textit{Exclusive} for a lengthy period. 
The loop at $\textit{LOE2}$ occurs when another thread is in $\textit{Shared}$ and refuses to leave. 
The loop in $\textit{LeaveE1}$ is required to obtain a concise equivalent external behaviour. 
When the busy protocol is used as intended, \ie, threads only use common accesses for a short time, 
and the implementation uses the right internal scheduling, these loops rarely occur. They are therefore marked 
\textit{improbable}.

\subsection{The implementation of the busy-forbidden protocol}
\label{section:busy_forbidden_implementation}
The code for 
entering and leaving the exclusive sections is described in Table~\ref{table:busy_forbidden}.
The busy-forbidden protocol is implemented by assigning to each thread two atomic flags,
called $\textit{busy}$ and $\textit{forbidden}$. The flag $\textit{busy}$ indicates that the current
thread is in its shared section and can only be written to by this thread. 
The flag $\textit{forbidden}$ indicates that some thread is having exclusive access.

\begin{table}[]
\begin{center}
\begin{tabular}{|l|l|}
\hline
\resetlinenumber
\begin{minipage}{0.4\textwidth}
\resetlinenumber
\begin{internallinenumbers}
$\texttt{enter\_shared}(\textit{thread}~p)$\\
\hspace*{0.4cm}$p.\textit{busy}:=\true$;\\
\hspace*{0.4cm}$\textbf{while}~p.\textit{forbidden}$\\
\hspace*{0.8cm}$    p.\textit{busy}:=\false;$\\
\hspace*{0.8cm}$    \textbf{if}~\textit{mutex}.\textit{timed\_lock}()$\\
\hspace*{1.2cm}$\textit{mutex}.\textit{unlock}();$\\
\hspace*{0.8cm}$    p.\textit{busy}:=\true;$
\\
\end{internallinenumbers}
\end{minipage}&
\begin{minipage}{0.4\textwidth}
\resetlinenumber
\begin{internallinenumbers}
$\texttt{enter\_exclusive}(\textit{thread}~p)$\\
\hspace*{0.4cm}$\textit{mutex}.\textit{lock}();$\\
\hspace*{0.4cm}$\textbf{while exists}~\textit{thread}~q\textbf{ with}$\\
\hspace*{1.2cm}$\neg q.\textit{forbidden}$\\
\hspace*{0.8cm}$    \textbf{select}~\textit{thread}~r$\\
\hspace*{0.8cm}$      r.\textit{forbidden}:=\true;$\\
\hspace*{0.8cm}$    \textbf{if}~r.\textit{busy} ~\textbf{or sometimes}$\\
\hspace*{1.2cm}$        r.\textit{forbidden}:=\false;$
\end{internallinenumbers}
\end{minipage}\\
\hline
\begin{minipage}{0.4\textwidth}
\resetlinenumber
\begin{internallinenumbers}
$\texttt{leave\_shared}(\textit{thread}~p)$\\
\hspace*{0.4cm}$p.\textit{busy}:=\false;$
\\ \\ \\ \\ \\ \\ \\
\end{internallinenumbers}
\end{minipage}&
\begin{minipage}{0.4\textwidth}
\resetlinenumber
\begin{internallinenumbers}
$\texttt{leave\_exclusive}(\textit{thread}~p)$\\
\hspace*{0.4cm}$\textbf{while exists}\textit{ thread q}\textbf{ with}$\\
\hspace*{1.2cm}$q.\textit{forbidden}$\\
\hspace*{0.8cm}$\textbf{select }\textit{thread } r$\\
\hspace*{0.8cm}\textbf{usually do}\\
\hspace*{1.2cm}$ r.\textit{forbidden}:=\false;$\\
\hspace*{0.8cm}\textbf{sometimes do}\\
\hspace*{1.2cm}$ r.\textit{forbidden}:=\true$\\
\hspace*{0.4cm}$mutex.unlock();$
\end{internallinenumbers}
\end{minipage}\\
\hline
\end{tabular}
\end{center}
\caption{Pseudocode description of the busy-forbidden protocol.}
\label{table:busy_forbidden}
\end{table}

Besides the flags there
is one generic mutual exclusion variable, called \textit{mutex}. 
The variable
\textit{mutex} can not only be locked and unlocked, but also provides a timed lock operation
$\textit{timed\_lock}()$. It tries to lock the mutex, and if that fails after a certain
time, it returns false without locking it. The timed lock is only important
for performance, and can be replaced by a wait instruction or even be omitted altogether.

When entering the shared section, a thread
generally only accesses its own \textit{busy} and \textit{forbidden} flags as $\textit{forbidden}$ is 
almost always false. 
These flags are only rarely accessed by other threads and therefore virtually always
available in the local cache of the processor executing the thread. 
In the rare case when the $\textit{forbidden}$
flag is set, this thread backs off using $\textit{mutex}$ to try again later. 
In principle the while-loop can be iterated indefinitely, giving rise to the internal loop in 
state $\textit{EnterS}$ in the specification. 
Leaving the shared section consists of only setting the $\textit{busy}$ flag of the thread to false.

Accessing the exclusive section is far more expensive. By using \textit{mutex}, mutual
access to the exclusive section is obtained. 
Subsequently, the $\textit{forbidden}$ flag for each thread $p$ is set to true, unless the $\textit{busy}$ flag of thread $p$ is set, as in this case the $\textit{forbidden}$ flag must be set to false again.
 
There is a non immediately obvious scenario where one thread refuses to leave the shared section, 
and two other threads $p_2$ and $p_3$ want to access the shared and exclusive section, respectively. 
Thread $p_3$ cannot obtain exclusive access, but hence should not indefinitely block shared
access for $p_2$. Hence, $p_3$ must set the $\textit{forbidden}$ flag of $p_2$ to false
if $\textit{busy}$ of $p_1$ is true. 

Without the \textbf{sometimes} part, which represents an arbitrary heuristic which only rarely holds, the implementation is not divergence-preserving
bisimilar to the specification, as reading $r.\textit{busy}=\false$ in line 9, once all other forbidden flags have been set, leads to a state without an internal loop, which does not occur in the specification. 
Without the \textbf{sometimes} part, a matching specification would become substantially more complex exhibiting
exactly when each $\textit{forbidden}$ flag is set, rendering the specification far less abstract and hence making it less useful.  

When leaving the exclusive section a thread resets all \textit{forbidden}
flags of the other threads. If this is done in a predetermined sequence the divergence-preserving 
branching bisimilar external behaviour becomes very complex, as this sequence has an influence on
the precise sequence other threads can enter the shared section. By resetting and sometimes even
setting the \textit{forbidden} flag, a comprehensible provably equal external behaviour is obtained,
although it leads to another $\tau$-loop in the specification. 
Practically, re-resetting the flag is hardly ever needed, certainly not for the Term Library. 
However, it is interesting to further investigate the optimal use of the timing of \textit{mutex} in 
\texttt{enter\_shared},
as well as the optimal rate of occurrence of the \textbf{sometimes} instructions for generic
uses of the busy-forbidden protocol. 

We modelled the specification and implementation of the busy-forbidden
protocol in mCRL2 (see Section \ref{section:busy_forbidden_verification}) and proved them 
divergence-preserving branching bisimilar.

\subsection{Behavioural properties of the busy-forbidden protocol}\label{section:busy_forbidden_correctness}

As an extra check we also formulate a number of natural requirements that
should hold for this protocol. These requirements have been verified by formulating them as modal properties. 
\begin{enumerate}
    \item \label{item:prop_mutex_exclusive} There should never be
    more than one thread present in the exclusive section.
    \item \label{item:prop_crtical_mutex_w_shared} There should never be a thread present in the exclusive section while one or more threads are present in the shared section.
    \item \label{item:prop_event_shared}
          When a thread requests to enter the shared section, it will be
          granted access within a bounded number of steps, unless there is another thread in the exclusive section. 
    \item \label{item:prop_event_crit}
        When a thread requests to enter the exclusive section, it will be granted access within a bounded number of steps, unless there is another thread in the shared or in the exclusive section.
    \item \label{item:prop_leave}
          When a thread requests to leave the exclusive/shared section, it will
          leave it within a bounded number of steps. 
    \item \label{item:prop_enter} A thread not in the exclusive or shared section can instantly start to 
          enter the exclusive or shared section. 
\end{enumerate}
For properties \ref{item:prop_event_shared}, \ref{item:prop_event_crit},  and \ref{item:prop_leave} granting access
and leaving can be indefinitely postponed if other threads are entering and leaving
exclusive and shared sections, or when other threads are in the while loops, continuously
writing forbidden and busy flags. This means that the algorithm relies on fair scheduling
of threads. 

\subsection{Existing readers-writer locks}
Common readers-writer locks, such as $\texttt{std::shared\_mutex}$ in $\cpp$17 in MSVC, use a mutual exclusion variable when entering and leaving the shared/reader section. 
This leads to poor scalability and is one of the reasons why the usage of readers-writer locks is often discouraged \cite{CBBJ08}.

The readers-writer lock by Mellor-Crumney and Scott \cite{MCS91} reduces resource contention by using a counter to keep track of the amount of current readers and introducing a queue system in which threads only have to notify the thread next in line when they leave the lock.
This lock is further improved by Krieger et al.\  \cite{Krieger93} by reducing the amount of shared variables to a single pointer and using a double-linked list instead of a queue such that reader threads can leave the lock without having to wait for neighbouring readers to also be done reading. However, the single shared pointer needs to be updated using a costly compare-and-swap operation every time a thread enters the lock, which becomes a bottleneck when multiple threads try to enter at the same time.

Lev et al.\ \cite{Lev09} provide several readers-writer lock algorithms aimed at improving scalability by significantly reducing resource 
contention through the use of a tree-like data structure called C-SNZI. Lev et al.\ show that their algorithms outperform other readers-writer 
locks when the majority of accesses, \ie, $ \geq 80\%$, are read accesses. Concurrent read-accesses however still have resource contention 
with a $100\%$ read workload, whereas the busy-forbidden protocol has none. Thus, this implies that the busy-forbidden protocol outperforms the C-SNZI based algorithms for our use case.

\subsection{Performance of the busy-forbidden protocol}
We have implemented the busy-forbidden protocol in $\cpp$ and assess its scalability compared to that of the $\texttt{std::shared\_mutex}$ by having threads repeatedly enter and leave the shared/reader section or the exclusive/writer section. Each thread uses a random number generator to decide on which section to enter and then leave, with a preset probability of $99.99\%$ of a thread deciding to enter and leave the shared/reader section. This probability was chosen as it corresponds to that of a typical use case such as state space generation. 

\begin{wrapfigure}{r}{7cm}
\hspace{6ex}
\centering
\begin{tikzpicture}
\begin{axis}[
  legend style = {at={(0.5,1.1)}, anchor=south},
  xtick={1,2,3,4,6,8},
  xmin = 0,
  xmax = 9,
  xlabel={\#threads},
  ymin = 0,
  ymax = 100000,
  ylabel={wallclock time (ms)},
  width=7cm,
  height=4cm
  ]
  \addplot[mark options={fill=gray}, mark=triangle*] coordinates {   
(1, 1660) (2,2239) (3,3929) (4,5564) (6,8977) (8,11823) 
  };
  \addlegendentry{busy-forbidden}
  \addlegendentry{std::shared\_mutex}
  \addplot[mark options={fill=gray}, mark=square*] coordinates {(1,12488) (2,46187) (3,57000) (4,85857)};
\end{axis}
\end{tikzpicture}
\caption{readers-writer lock benchmarks.}
\label{fig:scalability-busyforbidden}
\end{wrapfigure}
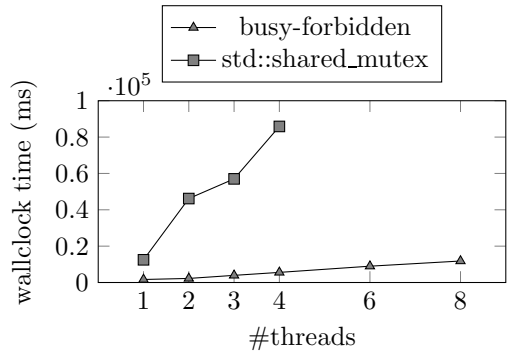

Figure \ref{fig:scalability-busyforbidden} shows the wall clock time of $\#\textit{threads}$ threads entering and then 
leaving a random section $ 10^9 $ times per thread.
The values displayed are the averages of 5 different runs.
The measurements were taken on an Intel i7-7700HQ processor and the $\cpp$ code was compiled using the MSVC19 compiler 
with the $\texttt{-O2}$ flag enabled.
The wall clock time of $\texttt{std::shared\_mutex}$ for more than $4$ threads is omitted from the graph as it is too large to nicely display.

We observe in Figure \ref{fig:scalability-busyforbidden} that the busy-forbidden protocol performs significantly better than the $\texttt{std::shared\_mutex}$ and costs only a minimal amount of overhead. This can also be seen in Figure \ref{fig::addbenchmarks} (a), 
discussed in Section \ref{section:performance_eval}, in which the busy-forbidden protocol and $\texttt{std::shared\_mutex}$ 
are used in combination with our implementation of the parallel Term Library.

\section{Modelling and verifying the algorithms}\label{section:busy_forbidden_verification}
As parallel algorithms are hard to get correct, we made models of the busy-forbidden protocol, of both specification and implementation, and the thread-safe Term Library in the process modelling language mCRL2 and verified
the properties by formulating them in the modal mu-calculus  \cite{groote2014modeling}. 
The specification model is a direct reflection of the external behaviour shown in Figure \ref{fig:external-busy-forbidden}.
The resulting implementation models are a direct reflection of the pseudocode in Table \ref{table:par_aterm} and 
\ref{table:busy_forbidden}. The formulas are a one to one translation of the requirements listed
in this article. For this reason, and for the reason of space, the models and formulas, are not
included in this article\footnote{All models and formulas
can be found in Appendices \ref{app:busy-forbidden} and \ref{app:aterm-impl}, respectively.}. 

Due to the nature of model checking, we only verify the models for finite instances. 
We repeatedly found that when protocols or distributed systems are erroneous, the problems
already reveal themselves in small instances \cite{DBLP:conf/forte/GrooteK21}. 
Used in this way, model checking is so efficient that it can effectively be used within the workflow
of constructing software. 
The busy-forbidden protocol was modelled and proven, before implementation commenced, and we did not run 
into any problem with it during implementation. 

A general equivalence proof has since been given for the specification and implementation of the busy-forbidden protocol \cite{phmvsp22}, using an extension of the Cones and Foci method\cite{DBLP:journals/jlp/GrooteS01,DBLP:journals/fmsd/FokkinkPP06}.

The correctness of the protocol and library have not been proven in general for any number of threads and terms. 
Unfortunately, we do not know of any effective method to prove modal formulas on models with 
a complexity such as ours, either automatically or manually, for any number of threads and terms,
and consider this an important direction of research. 

The model of the busy-forbidden protocol does not include the $\textit{mutex}.\textit{timed\_-}$ $\textit{lock}()$ statement as it is only important for performance.
The \textbf{sometimes} keywords are modelled as non-deterministic choices.
The specification and implementation are proven to be divergence 
preserving branching bisimulation equivalent, for up to 7 concurrent threads. 

We transformed the six requirements discussed in Section \ref{section:busy_forbidden_correctness} into modal logic formulas, and verified them both on the specification and the implementation,
although the latter was not really necessary due to their equivalency.
The equivalence and properties were verified, both on the specification as well as on the implementation model, for up to 7 threads. 
We uncovered a number of issues and obtained various insights while doing the verification for 2 and 3 threads. 
The verification with more threads, although increasingly time consuming, did not lead to
any additional insight. 

The model is primarily concerned with the thread-safe creation and garbage collection of terms, and therefore the typical term structure, where terms contain subterms, is also not part of the model, and as such only uses terms with arity $0$.
We use the equivalent specification model of the busy-forbidden protocol instead of its implementation model, as it is significantly smaller.
Furthermore, the model of the Term Library does not include buckets or a hashing function. 
Instead the hash table is modelled as a simple associative array, with atomic \textit{contains} and 
\textit{insert} operations.

The four properties discussed in Section \ref{behaviouralproperties} are also translated into modal logic and verified for finite instances. 
We have verified these properties for up to $3$ threads, using $3$ different terms and $4$ possible addresses, 
giving us reasonable certainty that the thread-safe Term Library works as intended. 
We were unable to verify our properties on larger state spaces as they became too big to verify automatically.
For example the state space of the aforementioned setup with 4 threads instead of 3 has 129 billion states.

\section{Performance evaluation} \label{section:performance_eval}

We have implemented a sequential and a thread-safe version of the Term Library. 
Both of these implementations are almost identical except for the synchronisation primitives added to 
the thread-safe version where necessary, including the busy-forbidden protocol.
Furthermore, we have implemented both reference counting and address protection sets as garbage 
collection strategies in both implementations for comparison. We compare these with the 
sequential ATerm library as used in the mCRL2 toolset \cite{groote2014modeling} and with a thread-safe Java implementation \cite{LankampParAterm} of the Term Library used in tools such as Spoofax \cite{KV10}, which was the only other thread-safe term library that we could find.
All reported measurements are the average of five runs with an AMD EPYC 7452 32-Core processor, unless stated otherwise.

\begin{figure}[H]
\centering
\begin{subfigure}[t]{0.49\textwidth}
\begin{tikzpicture}
\begin{axis}[
  xtick={0,4,8,12,16,20,24,28,32},
  xmin = .8,
  xmax = 32.2,
  ymin = 0,
  ymax = 2.2,
  width=7cm
  ]
  \addplot[mark options={fill=gray}, mark=triangle*] coordinates {   
(1, 0.03) (2, 0.11) (3, 0.22) (4, 0.14) (5, 0.26) (6, 0.34) (7, 0.28) (8, 0.29) (9, 0.39) (10, 0.51) (11, 0.53) (12, 0.51) (13, 0.59) (14, 0.5) (15, 0.54) (16, 0.49) (17, 0.53) (18, 0.54) (19, 0.48) (20, 0.5) (21, 0.48) (22, 0.52) (23, 0.5) (24, 0.53) (25, 0.5) (26, 0.49) (27, 0.53) (28, 0.51) (29, 0.51) (30, 0.49) (31, 0.48) (32, 0.48)
  };
  \addplot[mark options={fill=gray}, mark=square*] coordinates {
(1, 0.03) (2, 0.07) (3, 0.07) (4, 0.2) (5, 0.15) (6, 0.28) (7, 0.2) (8, 0.17) (9, 0.32) (10, 0.34) (11, 0.32) (12, 0.35) (13, 0.39) (14, 0.32) (15, 0.39) (16, 0.33) (17, 0.41) (18, 0.39) (19, 0.38) (20, 0.38) (21, 0.37) (22, 0.39) (23, 0.43) (24, 0.37) (25, 0.39) (26, 0.41) (27, 0.39) (28, 0.39) (29, 0.42) (30, 0.44) (31, 0.39) (32, 0.41)  
  };
  \addplot[mark options={fill=gray}, mark=*] coordinates {
  (1, 0.02) (32, 0.02)
  };
  \addplot[mark options={fill=gray}, mark=diamond*] coordinates {
  (1, 0.02) (32, 0.02)
  };
  \addplot[mark options={fill=gray}, mark=+] coordinates {
  (1, 0.01) (32, 0.01)
  };
  \addplot[dashed] coordinates {
(1, 0.26) (2, 0.46) (3, 0.68) (4, 1.32) (5, 1.25) (6, 1.2) (7, 1.51) (8, 1.41) (9, 1.36) (10, 1.48) (11, 1.51) (12, 1.4) (13, 1.51) (14, 1.44) (15, 1.43) (16, 1.38) (17, 1.67) (18, 1.77) (19, 1.87) (20, 1.77) (21, 1.78) (22, 1.85) (23, 1.68) (24, 1.95) (25, 1.68) (26, 1.82) (27, 1.81) (28, 1.65) (29, 1.94) (30, 1.94) (31, 2.09) (32, 1.86) 
  };
  \addplot[dotted] coordinates {
  (1, 0.03) (2, 0.12) (3, 0.18) (4, 0.32) (5, 0.24) (6, 0.22) (7, 0.38) (8, 0.47) (9, 0.47) (10, 0.55) (11, 0.5) (12, 0.58) (13, 0.62) (14, 0.63) (15, 0.67) (16, 0.72) (17, 0.74) (18, 0.76) (19, 0.79) (20, 0.83) (21, 0.83) (22, 0.88) (23, 0.91) (24, 0.95) (25, 0.96) (26, 0.99) (27, 0.98) (28, 1.04) (29, 1.02) (30, 1.1) (31, 1.11) (32, 1.16) 
  };
\end{axis}
\end{tikzpicture}
\caption{Creating new terms (shared).}\label{fig:results_shared_create}
\end{subfigure}\hfill
\begin{subfigure}[t]{0.49\textwidth}
\pgfplotsset{scaled y ticks=false}
\begin{tikzpicture}
\begin{axis}[
  xtick={0,4,8,12,16,20,24,28,32},
  xmin = .8,
  xmax = 32.2,
  ymin = 0,
  ymax = 0.5,
  width=7cm
  ]
  \addplot[mark options={fill=gray}, mark=triangle*] coordinates {
  (1, 0.03) (2, 0.03) (3, 0.03) (4, 0.04) (5, 0.03) (6, 0.04) (7, 0.05) (8, 0.07) (9, 0.06) (10, 0.06) (11, 0.06) (12, 0.06) (13, 0.05) (14, 0.05) (15, 0.07) (16, 0.06) (17, 0.06) (18, 0.05) (19, 0.06) (20, 0.07) (21, 0.06) (22, 0.06) (23, 0.07) (24, 0.06) (25, 0.05) (26, 0.06) (27, 0.06) (28, 0.07) (29, 0.07) (30, 0.07) (31, 0.06) (32, 0.06)   
  };    
  \addplot[mark options={fill=gray}, mark=square*] coordinates {
  (1, 0.03) (2, 0.05) (3, 0.03) (4, 0.04) (5, 0.03) (6, 0.03) (7, 0.04) (8, 0.05) (9, 0.05) (10, 0.05) (11, 0.06) (12, 0.05) (13, 0.06) (14, 0.06) (15, 0.05) (16, 0.06) (17, 0.07) (18, 0.06) (19, 0.05) (20, 0.06) (21, 0.06) (22, 0.06) (23, 0.06) (24, 0.06) (25, 0.05) (26, 0.07) (27, 0.06) (28, 0.06) (29, 0.05) (30, 0.05) (31, 0.05) (32, 0.06) 
  };
  \addplot[mark options={fill=gray}, mark=*] coordinates {
  (1, 0.02) (32, 0.02)
  };
  \addplot[mark options={fill=gray}, mark=diamond*] coordinates {
  (1, 0.02) (32, 0.02)
  };
  \addplot[mark options={fill=gray}, mark=+] coordinates {
  (1, 0.01) (32, 0.01)
  };
\addplot[dashed] coordinates {
(1, 0.26) (2, 0.25) (3, 0.28) (4, 0.26) (5, 0.28) (6, 0.28) (7, 0.28) (8, 0.35) (9, 0.33) (10, 0.35) (11, 0.32) (12, 0.33) (13, 0.35) (14, 0.32) (15, 0.34) (16, 0.33) (17, 0.34) (18, 0.33) (19, 0.35) (20, 0.33) (21, 0.34) (22, 0.34) (23, 0.37) (24, 0.34) (25, 0.35) (26, 0.35) (27, 0.37) (28, 0.35) (29, 0.34) (30, 0.34) (31, 0.35) (32, 0.34)
  };
  \addplot[dotted] coordinates {
  (1, 0.03) (2, 0.04) (3, 0.04) (4, 0.05) (5, 0.1) (6, 0.04) (7, 0.04) (8, 0.09) (9, 0.09) (10, 0.09) (11, 0.09) (12, 0.09) (13, 0.1) (14, 0.1) (15, 0.1) (16, 0.1) (17, 0.09) (18, 0.11) (19, 0.09) (20, 0.12) (21, 0.11) (22, 0.13) (23, 0.13) (24, 0.11) (25, 0.14) (26, 0.09) (27, 0.15) (28, 0.18) (29, 0.15) (30, 0.13) (31, 0.16) (32, 0.13)
  };
\end{axis}
\end{tikzpicture}
\caption{Creating new terms (distinct).}\label{fig:results_unique_create}
\end{subfigure}\hfill

\begin{subfigure}[t]{0.49\textwidth}
\begin{tikzpicture}
\begin{axis}[
  xtick={0,4,8,12,16,20,24,28,32},
  xmin = .8,
  xmax = 32.2,
  ymin = 0,
  ymax = 75,
  width=7cm
  ]
  \addplot[mark options={fill=gray}, mark=triangle*] coordinates {
  (1, 9.01) (2, 35.09) (3, 11.45) (4, 19.01) (5, 9.81) (6, 10.57) (7, 6.4) (8, 6.0) (9, 4.9) (10, 4.85) (11, 3.88) (12, 3.51) (13, 3.1) (14, 2.86) (15, 2.77) (16, 2.66) (17, 2.45) (18, 2.61) (19, 2.43) (20, 2.33) (21, 2.27) (22, 2.17) (23, 2.16) (24, 2.1) (25, 2.08) (26, 2.04) (27, 2.03) (28, 1.97) (29, 1.82) (30, 1.87) (31, 1.81) (32, 1.81)
  };
  \addplot[mark options={fill=gray}, mark=square*] coordinates {
  (1, 4.07) (2, 2.51) (3, 1.66) (4, 1.39) (5, 1.07) (6, 0.96) (7, 0.81) (8, 0.75) (9, 0.67) (10, 0.59) (11, 0.53) (12, 0.49) (13, 0.46) (14, 0.43) (15, 0.41) (16, 0.4) (17, 0.39) (18, 0.37) (19, 0.36) (20, 0.35) (21, 0.32) (22, 0.31) (23, 0.32) (24, 0.31) (25, 0.29) (26, 0.3) (27, 0.28) (28, 0.27) (29, 0.29) (30, 0.28) (31, 0.28) (32, 0.29)
  };
  \addplot[mark options={fill=gray}, mark=*] coordinates {
  (1, 4.01) (32, 4.01)
  };
  \addplot[mark options={fill=gray}, mark=diamond*] coordinates {
  (1, 3.65) (32, 3.65)
  };
  \addplot[mark options={fill=gray}, mark=+] coordinates {
  (1, 4.57) (32, 4.57)
  };
  \addplot[dashed] coordinates {
(1, 104.6) (2, 135.57) (3, 106.18) (4, 103.47) (5, 91.7) (6, 84.16) (7, 75.47) (8, 71.1) (9, 64.47) (10, 61.59) (11, 57.5) (12, 54.33) (13, 51.91) (14, 49.4) (15, 48.72) (16, 46.51) (17, 47.73) (18, 47.85) (19, 48.11) (20, 48.63) (21, 47.68) (22, 46.57) (23, 45.87) (24, 45.44) (25, 45.76) (26, 44.89) (27, 44.76) (28, 42.38) (29, 42.41) (30, 42.85) (31, 42.14) (32, 42.05)
  };
  \addplot[dotted] coordinates {
  (1, 6.51) (2, 14.21) (3, 15.1) (4, 15.11) (5, 18.7) (6, 20.71) (7, 22.04) (8, 33.32) (9, 28.07) (10, 25.38) (11, 24.46) (12, 22.8) (13, 22.65) (14, 23.12) (15, 22.64) (16, 22.47) (17, 22.93) (18, 23.23) (19, 23.67) (20, 24.74) (21, 24.06) (22, 24.3) (23, 24.65) (24, 24.35) (25, 24.96) (26, 25.05) (27, 25.4) (28, 25.51) (29, 26.01) (30, 26.69) (31, 27.46) (32, 28.26) 
  };
\end{axis}
\end{tikzpicture}
\caption{Creating existing terms (shared).}\label{fig:results_shared_lookup}
\end{subfigure}
\begin{subfigure}[t]{0.49\textwidth}
\begin{tikzpicture}
\begin{axis}[
  xtick={0,4,8,12,16,20,24,28,32},
  xmin = .8,
  xmax = 32.2,
  ymin = 0,
  ymax = 35,
  width=7cm
  ]
  \addplot[mark options={fill=gray}, mark=triangle*] coordinates {
  (1, 9.02) (2, 5.15) (3, 3.95) (4, 3.05) (5, 2.64) (6, 2.25) (7, 2.38) (8, 2.41) (9, 2.42) (10, 2.49) (11, 2.24) (12, 2.22) (13, 2.27) (14, 2.26) (15, 2.2) (16, 2.39) (17, 2.55) (18, 2.63) (19, 2.58) (20, 2.72) (21, 2.67) (22, 2.67) (23, 2.69) (24, 2.69) (25, 2.77) (26, 2.76) (27, 2.8) (28, 2.77) (29, 2.82) (30, 2.84) (31, 2.86) (32, 2.92) 
  };  
  \addplot[mark options={fill=gray}, mark=square*] coordinates {
  (1, 3.96) (2, 2.66) (3, 2.58) (4, 2.44) (5, 2.27) (6, 1.76) (7, 1.92) (8, 1.86) (9, 1.95) (10, 1.91) (11, 1.79) (12, 1.78) (13, 1.76) (14, 1.77) (15, 1.75) (16, 1.81) (17, 1.82) (18, 1.97) (19, 2.05) (20, 2.04) (21, 2.07) (22, 2.07) (23, 2.07) (24, 2.1) (25, 2.15) (26, 2.05) (27, 2.12) (28, 2.11) (29, 2.17) (30, 2.22) (31, 2.27) (32, 2.22)
  };
  \addplot[mark options={fill=gray}, mark=*] coordinates {
  (1, 4.02) (32, 4.02)
  };
  \addplot[mark options={fill=gray}, mark=diamond*] coordinates {
  (1, 3.71) (32, 3.71)
  };
  \addplot[mark options={fill=gray}, mark=+] coordinates {
  (1, 4.58) (32, 4.58)
  };
  \addplot[dotted] coordinates {
  (1, 6.46) (2, 13.79) (3, 15.45) (4, 15.74) (5, 18.34) (6, 18.53) (7, 24.56) (8, 33.17) (9, 26.9) (10, 25.04) (11, 23.73) (12, 22.7) (13, 22.35) (14, 22.93) (15, 22.58) (16, 22.24) (17, 22.72) (18, 23.19) (19, 23.81) (20, 24.65) (21, 24.05) (22, 24.14) (23, 24.4) (24, 24.4) (25, 25.29) (26, 25.31) (27, 25.93) (28, 25.7) (29, 26.28) (30, 27.09) (31, 27.79) (32, 28.69)
  };
\end{axis}
\end{tikzpicture}
\caption{Creating existing terms (distinct).}\label{fig:results_unique_lookup}
\end{subfigure}

\begin{subfigure}[t]{0.49\textwidth}
\begin{tikzpicture}
\begin{axis}[
  xtick={0,4,8,12,16,20,24,28,32},
  xmin = .8,
  xmax = 32.2,
  ymin = 0,
  width=7cm
  ]
  \addplot[mark options={fill=gray}, mark=triangle*] coordinates {       
   (1, 15.69) (2, 8.63) (3, 5.93) (4, 4.6) (5, 3.87) (6, 3.41) (7, 3.0) (8, 2.79) (9, 2.5) (10, 2.45) (11, 2.21) (12, 2.17) (13, 2.21) (14, 2.23) (15, 2.32) (16, 2.24) (17, 2.14) (18, 2.3) (19, 2.21) (20, 2.11) (21, 2.21) (22, 2.09) (23, 2.17) (24, 2.18) (25, 2.13) (26, 2.07) (27, 2.06) (28, 2.09) (29, 2.06) (30, 2.05) (31, 2.13) (32, 2.1)
  
  };
  \addplot[mark options={fill=gray}, mark=square*] coordinates {
  (1, 16.65) (2, 8.9) (3, 6.07) (4, 4.66) (5, 3.93) (6, 3.37) (7, 3.01) (8, 2.8) (9, 2.55) (10, 2.41) (11, 2.34) (12, 2.28) (13, 2.21) (14, 2.3) (15, 2.35) (16, 2.21) (17, 2.26) (18, 2.35) (19, 2.25) (20, 2.33) (21, 2.28) (22, 2.21) (23, 2.26) (24, 2.15) (25, 2.12) (26, 2.16) (27, 2.13) (28, 2.07) (29, 2.09) (30, 2.16) (31, 2.03) (32, 2.03) 
  
  };
  \addplot[mark options={fill=gray}, mark=*] coordinates {
  (1, 16.78) (32, 16.78)
  };
  \addplot[mark options={fill=gray}, mark=diamond*] coordinates {
  (1, 18.18) (32, 18.18)
  };
  \addplot[mark options={fill=gray}, mark=+] coordinates {
  (1, 16.36) (32, 16.36) 
  };
  \addplot[dashed] coordinates {
(1, 34.58) (2, 34.54) (3, 36.02) (4, 36.7) (5, 36.07) (6, 33.58) (7, 30.88) (8, 28.44) (9, 26.38) (10, 24.98) (11, 22.92) (12, 17.8) (13, 20.85) (14, 17.68) (15, 19.06) (16, 22.29) (17, 19.5) (18, 19.55) (19, 20.38) (20, 21.89) (21, 21.61) (22, 21.46) (23, 17.62) (24, 19.23) (25, 18.4) (26, 20.62) (27, 22.72) (28, 20.77) (29, 18.51) (30, 22.46) (31, 23.6) (32, 19.44)  
  };
  \addplot[dotted] coordinates {
  (1, 16.24) (2, 8.71) (3, 5.95) (4, 4.54) (5, 3.86) (6, 3.34) (7, 3.01) (8, 2.74) (9, 2.53) (10, 2.4) (11, 2.29) (12, 2.25) (13, 2.33) (14, 2.28) (15, 2.24) (16, 2.21) (17, 2.22) (18, 2.29) (19, 2.15) (20, 2.24) (21, 2.04) (22, 2.24) (23, 2.24) (24, 2.12) (25, 2.18) (26, 2.05) (27, 2.05) (28, 2.2) (29, 2.16) (30, 2.17) (31, 2.02) (32, 2.07) 
  };
\end{axis}
\end{tikzpicture}
\caption{Traversing terms (shared).}\label{fig:results_shared_inspect}
\end{subfigure}
\begin{subfigure}[t]{0.49\textwidth}
\begin{tikzpicture}
\begin{axis}[
  xtick={0,4,8,12,16,20,24,28,32},
  xmin = .8,
  xmax = 32.2,
  ymin = 0,
  ymax = 1,
  width=7cm,
  axis line style={draw=none},
  tick style={draw=none},
  legend cell align=left,
  legend style={at={(0.06,0.55)},anchor=west},
  yticklabels={,,},
  xticklabels={,,}
  ]
    
  \addplot[mark options={fill=gray}, mark=triangle*] coordinates {(4,0.5)}; 
  \addplot[mark options={fill=gray}, mark=square*] coordinates {(4,0.5)}; 
  \addplot[dashed] coordinates {(4,0.5)}; 
  \addplot[dotted] coordinates {(4,0.5)}; 
      
  \addplot[mark options={fill=gray}, mark=*] coordinates {(4,0.5)}; 
  \addplot[mark options={fill=gray}, mark=diamond*] coordinates {(4,0.5)}; 
  \addplot[mark options={fill=gray}, mark=+] coordinates {(4,0.5)};
    
  \addlegendentry[text width=4.5cm,text depth=]{thread-safe (busy-forbidden, reference count)}
  \addlegendentry[text width=4.5cm,text depth=]{thread-safe (busy-forbidden, protection set)}
  \addlegendentry{thread-safe (Java~\cite{LankampParAterm})}
  \addlegendentry{thread-safe (\textsf{std::shared\_mutex})}
  \addlegendentry{sequential (reference count)}
  \addlegendentry{sequential (protection set)}
  \addlegendentry{sequential (original)}
  
  \addlegendimage{empty legend}
  \addlegendentry{}
  \addlegendimage{empty legend}
  \addlegendentry{}
\end{axis}
\end{tikzpicture}
\caption{Legend for all plots in Figures~\ref{fig:results} and~\ref{fig::addbenchmarks}.}
\end{subfigure}
\caption{Execution time (in seconds) plotted against number of threads.}\label{fig:results}
\end{figure}
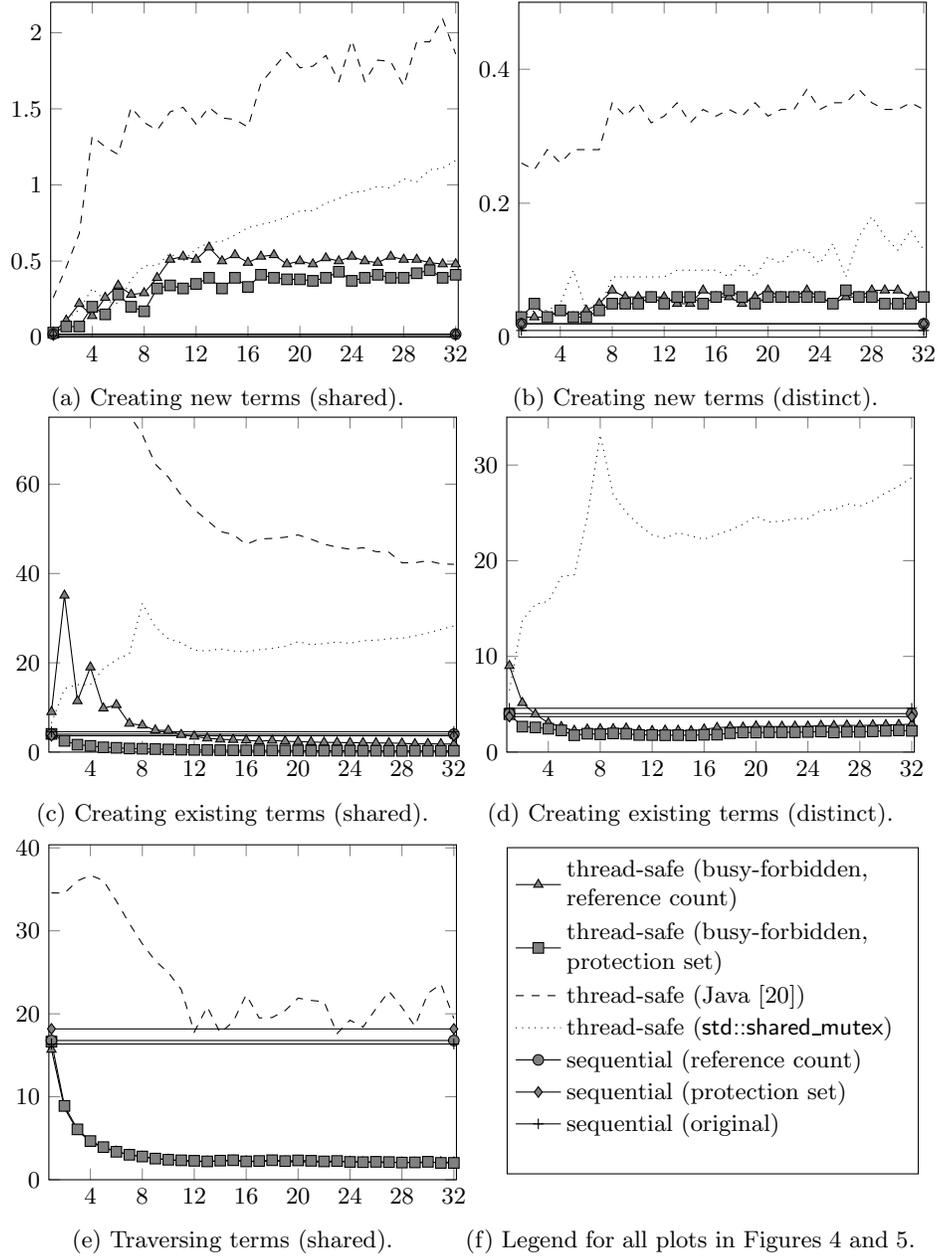

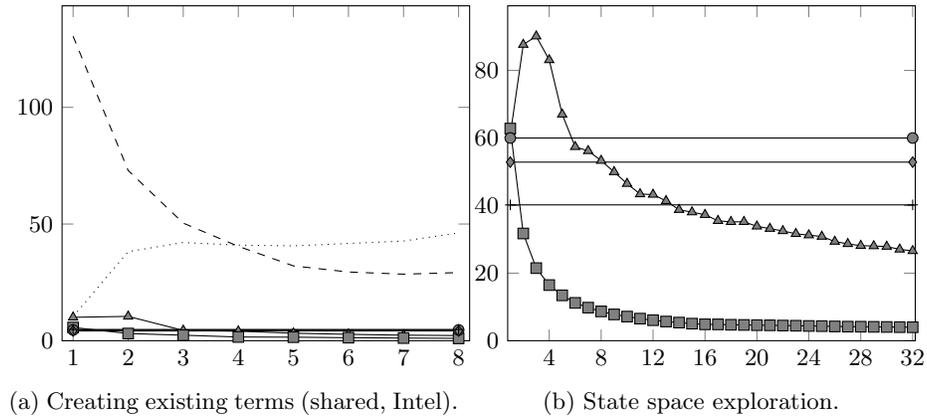
\begin{figure}
\begin{subfigure}[t]{0.49\textwidth}
\begin{tikzpicture}
\begin{axis}[
  xtick={0,1,2,3,4,5,6,7,8},
  xmin = .8,
  xmax = 8.2,
  ymin = 0,
  width=7cm
  ]
  \addplot[mark options={fill=gray}, mark=triangle*] coordinates {
  (1, 10.03) (2, 10.43) (3, 4.42) (4, 4.08) (5, 3.22) (6, 2.74) (7, 2.38) (8, 2.22)
  };
  \addplot[mark options={fill=gray}, mark=square*] coordinates {
  (1, 5.68) (2, 3.09) (3, 2.36) (4, 1.6) (5, 1.52) (6, 1.3) (7, 1.14) (8, 1.02) 
  };
  \addplot[mark options={fill=gray}, mark=*] coordinates {
  (1, 4.61) (8, 4.61)
  };
  \addplot[mark options={fill=gray}, mark=diamond*] coordinates {
  (1, 4.2) (8, 4.2)
  };
  \addplot[mark options={fill=gray}, mark=+] coordinates {
  (1, 4.83) (8, 4.83)
  };
  \addplot[dashed] coordinates {
  (1, 130.46) (2, 72.95) (3, 50.39) (4, 40.45) (5, 32.07) (6, 29.45) (7, 28.52) (8, 29.2)
  };
  \addplot[dotted] coordinates {
  (1, 10.34) (2, 38.17) (3, 42.09) (4, 40.96) (5, 40.65) (6, 41.67) (7, 42.72) (8, 46.19) 
  };
\end{axis}
\end{tikzpicture}
\caption{Creating existing terms (shared, Intel).}
\end{subfigure}
\begin{subfigure}[t]{0.49\textwidth}
\begin{tikzpicture}
\begin{axis}[
  xtick={0,4,8,12,16,20,24,28,32},
  xmin = .8,
  xmax = 32.2,
  ymin = 0,
  width=7cm
  ]
  \addplot[mark options={fill=gray}, mark=triangle*] coordinates {
  (1, 61.02) (2, 87.64) (3, 90.12) (4, 83.1) (5, 67.0) (6, 57.39) (7, 56.19) (8, 53.3) (9, 49.95) (10, 46.48) (11, 43.41) (12, 43.27) (13, 41.38) (14, 38.77) (15, 38.06) (16, 37.28) (17, 35.5) (18, 35.19) (19, 35.2) (20, 33.86) (21, 33.18) (22, 32.49) (23, 31.65) (24, 31.25) (25, 30.84) (26, 29.36) (27, 28.67) (28, 28.13) (29, 28.03) (30, 27.88) (31, 27.04) (32, 26.59)
  };
  \addplot[mark options={fill=gray}, mark=square*] coordinates {
  (1, 62.8) (2, 31.74) (3, 21.49) (4, 16.47) (5, 13.43) (6, 11.2) (7, 9.79) (8, 8.67) (9, 7.78) (10, 7.15) (11, 6.54) (12, 6.07) (13, 5.68) (14, 5.37) (15, 5.06) (16, 4.85) (17, 4.83) (18, 4.72) (19, 4.67) (20, 4.6) (21, 4.56) (22, 4.49) (23, 4.44) (24, 4.37) (25, 4.32) (26, 4.23) (27, 4.17) (28, 4.15) (29, 4.11) (30, 4.07) (31, 4.02) (32, 3.99)
  };
  \addplot[mark options={fill=gray}, mark=*] coordinates {
  (1, 60) (32, 60)
  };
  \addplot[mark options={fill=gray}, mark=diamond*] coordinates {
  (1, 52.86) (32, 52.86)
  };
  \addplot[mark options={fill=gray}, mark=+] coordinates {
  (1, 40.21) (32, 40.21)
  };
  \addplot[dashed] coordinates {
  
  };
  \addplot[dotted] coordinates {
  
  };
\end{axis}
\end{tikzpicture}
\caption{State space exploration.}
\end{subfigure}
\caption{Additional experiments comparing execution times versus threads.}\label{fig::addbenchmarks}
\end{figure}

The results are listed in the plots in Figures~\ref{fig:results} and \ref{fig::addbenchmarks}. 
In these plots the $y$-axis indicates the wall clock time in seconds and the $x$-axis the number of threads ($\#\textit{threads}$).
The triangles are the thread-safe reference count implementation and the squares the thread-safe set protection 
implementation. For the sequential versions we have circles for the reference count version, diamonds for the protection set version and plusses for the original implementation.
The results for the sequential implementations are extended horizontally for easier comparison.
Finally, the dashed line indicates the thread-safe Java implementation and the dotted line is our thread-safe 
implementation where the busy-forbidden protocol has been replaced by a \texttt{std::shared\_mutex}.
This last implementation uses protection sets. 

In Figure~\ref{fig:results} we report three experiments, one per row, designed to obtain insight in how the new thread-safe library 
performs for specific tasks. In the left column all threads access the same
shared term, whereas in the right column each thread operates on its own term, but these distinct terms are
stored in common data structures
and accessed via the hash table common for all threads. 

In Figure~\ref{fig:results} (a) we measure how expensive it is to create a term in parallel.
The threads create a term $t_{400\,000}$ defined as follows.
The term $t_0$ is equal to a constant $c$ and $t_i$ is $f(t_{i-1}, t_{i-1})$ for a function symbol $f$ of arity two, which is the most common arity used in practice. Note that due to sharing,
this term consists of 400\,001 term nodes. 
In (b) each thread creates the term $t_{400\,000/\#\textit{threads}}$ instead. With each term starting with a unique constant per thread, creating a total of $400\,000 + \#\textit{threads}$ term nodes.

In Figures~\ref{fig:results} (c) and (d) we measure the time it takes to create $1000/\#\textit{threads}$ 
instances of the terms used in respectively (a) and (b). This measures the time to create terms that
are already present in the term library, and this essentially boils down to a hash table lookup. 
In diagram (d) the Java results are left out as Java consistently requires more than 100 seconds. 
In the lower diagram, \ie, (e), we measure the time to perform $1000 / \#\textit{threads}$ 
breadth-first traversals on a term $t_{20}$.

The traversals do not employ the shared structure, hence $2^{21}-1$ terms are visited per traversal.
We observe that for this benchmark there is no difference in timings between traversing the term $t_{20}$ and traversing a unique term per thread.

We conclude that our term library completely outperforms the Java implementation. For
creating terms, the \texttt{std::shared\_mutex} is slower. For traversing terms no locking is required, and 
therefore, no difference is observed. The dotted\\

\noindent line is hidden under the line with the 
boxes\footnote{All benchmark results are listed in Appendix C.}. Except for 
creating new terms, the term library clearly benefits from the extra processors, outperforming the
sequential libraries with two processors using protection sets. When creating new terms, 
scaling goes reasonably well when beyond 12 processors. 
We observe in Figure~\ref{fig:results} (c) that the reference counting implementation for a few threads is
unexpectedly inefficient. In order to understand this, we retried the experiments on an Intel i7-7700HQ
processor, reported in Figure \ref{fig::addbenchmarks} (a). Here, none of the anomalies occur, and notably,
Java even outperforms the \texttt{std::shared\_mutex} implementation with more than four threads. 
This is in line with our many other experiments that
compiler and processor have a large influence on such benchmarks.

The dedicated benchmarks are promising, but in order to get insight in the behaviour of the Term Library in 
practical situations, we incorporated the Term Library in the mCRL2 toolset and used it to generate the state space
of the 1394 firewire protocol \cite{luttik1394}. Essentially each thread picks an
unexplored state from a common state buffer, and using term rewriting, generates all states reachable from
this state, putting them back in the buffer. With protection sets, two threads are already sufficient to
outperform all sequential implementations, and scaling is very good, where with 16 threads, the state 
space is generated more than 12 times faster. Reference counting is clearly a less viable option,
which is most likely due to the fact that often the same terms, such as $\true$ and $\false$, 
are accessed when calculating next states, leading to atomically changing the same reference count often. 
Note that in this prototype, nothing has been done yet to optimise
thread access to the common state buffer, being simply protected by a mutex. 

\newpage

\bibliographystyle{llncs/splncs04}
\bibliography{bibliography}

\newpage
\appendix
\section{The mCRL2 Language and Modal Formulas}
The models are written in mCRL2. 
This is a modelling language based on CCS (Calculus of Communicating Processes)
\cite{DBLP:books/sp/Milner80} and ACP (Algebra of Communicating Processes) \cite{DBLP:books/daglib/0069083}. 
It is based on atomic actions. Every occurrence of an atomic action causes a state
change. Typically, calling a function, or returning from a 
function, setting or reading a global variable are modelled by atomic actions. 
Each action consists of a label and a possible set of data parameters, \eg, the $\textit{lock}(p)$ action has $p$ as a data parameter.
The tau or hidden action $\tau$ has a special status, as it is an action of which
the occurrence cannot be observed directly.

Actions can be sequentially composed using the dot (`$\cdot$') operator. Alternative
composition, where nondeterministically one of the options can be chosen, is denoted using a plus (`$+$').
The $\sum_{e{:}S}$ operator denotes the application of the (`$+$') operator over all elements of some set $S$.
The if-then-else is written as $c{\rightarrow}p{\diamond}q$ where $p$ is executed if $c$ is true,
otherwise the process $q$ takes place. 

Parallel composition is denoted by $\parallel$ and allows the actions of two processes to both interleave and occur simultaneously.
Using the \textbf{comm} and \textbf{allow} operators, we can enforce that only specific actions can and must occur simultaneously. 
For example, $\textbf{comm}(\{a,b\}{\to}c\}, \textbf{allow}(\{c,d,e\}, (d\cdot a) || (b\cdot e)))$ enforces that the actions $d$ and $e$ can not occur simultaneously, while $a$ and $b$ must occur simultaneously with any other action, signified with $c$.
As a result, $d\cdot c \cdot e$ is the only possible sequence of actions that can occur.

Recursive behaviour is denoted using equations, typically
of the form $X=p$, \eg, $X=a{\cdot}X$ is the process that can perform an infinite number
of $a$'s. Similar to actions, the process variables $X$ can contain data parameters.
A counter can thus be described as $C(n{:}\Nat)=\textit{up}{\cdot}X(n+1)$.
An important type of data parameter that we use is a function. 
For example, the process variable $Y(m : \Nat \rightarrow \mathbb{B})$ uses a mapping $m$ from natural numbers to booleans. The function update $m[ n \mapsto b]$ specifies that $m[ n \mapsto b](k)$ equals $b$ if $k= n$, and otherwise,
it equals $m(k)$.

The safety and liveness properties that we verify, are written in the modal mu-calculus.
These consist of conjunctions ($\wedge$), disjunctions ($\vee$), implications ($\rightarrow$), negations ($\neg$), quantification ($\exists,\forall$) and $\true$ and $\false$, each with their usual meaning. 
Besides this there is a modality $\langle a \rangle \phi$ 
that is valid if we can take an action $a$ after which $\phi$ holds.
Similarly, the modality $[a]\phi$ holds iff after every possible $a$ action $\phi$ holds.
The action $a$ inside these modalities can also consist of possibly multiple actions. This can be done through sequential composition $(\cdot)$, choice $(\cup)$, intersection $(\cap)$ and complement $(\overline{a})$. 
For example, the formula $\langle \overline{a} \cup \overline{b} \rangle \true$ only holds if we can do some action that is neither $a$ or $b$. 
The expression $\true$ in a modality represents the set of all actions. 
Using Kleene's star on a set of actions, all sequences over the action in this set are expressed.
An often occurring pattern is $[\true^*]\phi$ expressing that $\phi$ must hold in all states 
reachable via a sequence of actions. 

We can also write recursive formulas using the minimal fixed point operator $\mu X.\phi$ and 
the maximal fixed point operator $\nu X. \phi$. For example, the formula $\nu X.\langle a \rangle X$ expresses that we must be able to perform action 
$a$ after which the same formula still holds. Thus this formula only holds if we can perform an infinite 
amount of $a$ actions. The difference between a minimal and a maximal fixed
point is that iteration through the fixed point 
variable must be bounded in a minimal fixed point. 

A fixed point construction used in several properties is 
\[
  \nu X.\mu Y.([\overline{\textit{succes} \cup \textit{interrupt}}]Y \land [\textit{interrupt}]X \wedge \langle \true^* \sequential \textit{succes} \rangle \true)
\]
This says that an action $\textit{succes}$ will always occur within a finite 
amount of steps, unless an action $\textit{interrupt}$ continuously occurs. 
But even in that case $\textit{succes}$ must remain possible.
This construction is useful for properties in which we state that something must eventually happen given fair scheduling. 

The fixed point operators also allow us to pass on parameters in the same way we can do for process variables. 
This allows us, for example, to keep track of the number of times that a given action has occurred.
We discuss one such fixed point operator in Appendix~\ref{app:busy-forbidden}.

\begin{table}[]
\begin{tabular}{|p{\textwidth}|} \hline
$\textit{BF}(s : P\rightarrow S) =~\sum_{p{:}P} \sequential ($\\
\tab{}{1}$(s(p) \approx \textit{Free})$\\
\tab{$\rightarrow$}{2}$\texttt{enter\_shared\_call}(p) \sequential \textit{BF}(s[p\mapsto \textit{EnterS}])$\\
\tab{+}{1}$(s(p) \approx \textit{EnterS})$\\
\tab{$\rightarrow$}{2}$((\neg\exists_{p':P}.~s(p') \in \{\textit{LOS}, \textit{Exclusive}\}) \rightarrow \tau \sequential \textit{BF}(s[p\mapsto \textit{LOE1}])$\\
\tab{$\diamond$}{2}$\textit{improbable} \sequential \textit{BF}(s))$\\
\tab{+}{1}$(s(p) \approx \textit{LOE1})$\\
\tab{$\rightarrow$}{2}$\texttt{enter\_shared\_return}(p) \sequential \textit{BF}(s[p\mapsto \textit{Shared}])$\\
\tab{+}{1}$(s(p) \approx \textit{Shared})$\\
\tab{$\rightarrow$}{2}$\texttt{leave\_shared\_call}(p) \sequential \textit{BF}(s[p\mapsto \textit{LeaveS}])$\\
\tab{+}{1}$(s(p) \approx \textit{LeaveS})$\\
\tab{$\rightarrow$}{2}$\texttt{leave\_shared\_return}(p) \sequential \textit{BF}(s[p\mapsto \textit{Free}])$\\
\tab{+}{1}$(s(p) \approx \textit{Free})$\\
\tab{$\rightarrow$}{2}$\texttt{enter\_exclusive\_call}(p) \sequential \textit{BF}(s[p\mapsto \textit{EnterE}])$\\
\tab{+}{1}$(s(p) \approx \textit{EnterE} \wedge \neg\exists_{p' : P}.~s(p') \in \{\textit{LOE2}, \textit{LOS}, \textit{Exclusive}\})$\\
\tab{$\rightarrow$}{2}$\tau \sequential \textit{BF}(s[p\mapsto \textit{LOE2}])$\\
\tab{+}{1}$(s(p) \approx \textit{LOE2})$\\
\tab{$\rightarrow$}{2}$\textit{improbable} \sequential \textit{BF}(s)$\\
\tab{+}{1}$(s(p) \approx \textit{LOE2} \wedge \neg\exists_{p' : P}.~s(p') \in \{\textit{LOE1}, \textit{Shared}\})$\\
\tab{$\rightarrow$}{2}$\tau \sequential \textit{BF}(s[p\mapsto \textit{LOS}])$\\
\tab{+}{1}$(s(p) \approx \textit{LOS})$\\
\tab{$\rightarrow$}{2}$\texttt{enter\_exclusive\_return}(p) \sequential \textit{BF}(s[p\mapsto \textit{Exclusive}])$\\
\tab{+}{1}$(s(p) \approx \textit{Exclusive})$\\
\tab{$\rightarrow$}{2}$\texttt{leave\_exclusive\_call}(p) \sequential \textit{BF}(s[p\mapsto \textit{LeaveE1}])$\\
\tab{+}{1}$(s(p) \approx \textit{LeaveE1})$\\
\tab{$\rightarrow$}{2}$\textit{improbable} \sequential \textit{BF}(s)$\\
\tab{+}{1}$(s(p) \approx \textit{LeaveE1})$\\
\tab{$\rightarrow$}{2}$\tau \sequential \textit{BF}(s[p \mapsto \textit{LeaveE2}])$\\
\tab{+}{1}$(s(p) \approx \textit{LeaveE2})$\\
\tab{$\rightarrow$}{2}$\texttt{leave\_exclusive\_return}(p) \sequential \textit{BF}(s[p\mapsto \textit{Free}])~)$\\ \hline
\end{tabular}

\caption{Specification of the busy-forbidden protocol corresponding to Figure~\ref{fig:external-busy-forbidden}.}
\label{spec:busy_forbidden}
\end{table}

\section{mCRL2 Specifications for the Busy-Forbidden Protocol}\label{app:busy-forbidden}

In this section we give the formal mCRL2 specifications of the implementation and the external behaviour, \ie, the specification, of the busy-forbidden protocol that are used to perform the model and equivalence checking.
The process specification given in Table~\ref{spec:busy_forbidden} exactly matches the external behaviour shown in Figure~\ref{fig:external-busy-forbidden}.
We define $P$ to be the (finite) set of threads and we define $\var{S}$ to be a data set representing the set of states: 
\vspace{-0.2cm}
\[\begin{tabular}{l}
\var{S} = \{ \textit{Free}, \textit{EnterS}, \textit{LOE1}, \textit{Shared}, \textit{LeaveS}, \\
\hspace*{2cm}\textit{EnterE}, \textit{LOE2}, \textit{LOS}, \textit{Exclusive},
\textit{LeaveE1}, \textit{LeaveE2} \}
\end{tabular}\]
The mapping $s$ maps each thread to their current state.
Initially, $s(p) = \textit{Free}$ for all $p \in P$.
The conditions for performing transitions are the same as the conditions in the diagram of the external behaviour.

Observe that we use a typewriter font (for example \texttt{enter\_shared\_call}) to indicate visible actions and an italics font (for example $\textit{store}_p$) to indicate internal actions that will be hidden for divergence-preserving branching bisimulation reductions.

The mCRL2 specification of the implementation is separated per function.
Entering the shared section is specified in Table~\ref{spec:enter_shared} and leaving it in Table~\ref{spec:leave_shared}.
Entering the exclusive section is specified in Table~\ref{spec:enter_exclusive} and leaving it in Table~\ref{spec:leave_exclusive}.

Note that we use actions to model the assignments to variables. For example $\textit{store}_p(\textit{Busy}(p),\true,p)$ corresponds to the assignment of $\true$ to $\textit{p.busy}$ in the implementation pseudocode.
The process algebra has no global variables and we use an additional process and actions to read from and write to these variables.
For the atomic flags we introduce a struct $F$ that is defined below to declare a busy and a forbidden flag per thread.
\begin{equation*}
  \texttt{sort}~F = \texttt{struct}~\textit{Busy}(P)~|~\textit{Forbidden}(P)
\end{equation*}
Table~\ref{spec:busy_forbidden_comm} shows the behaviour of the \textit{Busy} and \textit{Forbidden} flags for every thread and the \textit{mutex} variable.
We model the `while' construction in the pseudocode by recursion and have added the \emph{improbable} action to ensure equivalence modulo divergence-preserving branching bisimulation.
When entering the exclusive section, we use a set \emph{forbidden} (and for leaving \emph{allowed}) to keep track of the threads whose forbidden flag have already been set to $\true$ ($\false$ when leaving).

Table~\ref{spec:busy_forbidden_thread} shows the specification for the behaviour of a thread.
Each thread repeatedly chooses (non-deterministically) to enter and leave either the shared or exclusive section. 
Finally, Table~\ref{spec:busy_forbidden_init} contains the complete mCRL2 specification of the various processes in a parallel composition and the necessary communication to deal with the atomic flags and mutex.

As the next step, we transformed the six requirements discussed in Section~\ref{section:busy_forbidden_correctness} into modal logic formulas, and verified them on the specification.
Note that these properties are preserved by divergence-preserving branching bisimulation, so verifying the implementation is not necessary. 
We discuss property~\ref{item:prop_crtical_mutex_w_shared} in detail as an illustration of what such formulas look like.
The informal description of the property reads:
\begin{enumerate}
  \item[2.
  ] There should never be a thread present in the exclusive section while one or more threads are present in the shared section.
\end{enumerate}
The corresponding modal formula is shown in Table~\ref{property:busy_forbidden_p2}.
We use a maximal fixpoint with two data parameters, namely $n_{shared}$ and $n_{exclusive}$, both initially $0$. 
The argument $n_{shared}$ indicates the number of threads present in the shared section, and $n_{exclusive}$ the number in the exclusive section.
At lines $2$ through $5$, we keep track of the amount of threads present in each section, 
updating the variables after each respective action.
At lines $6$ through $11$, we state that our variables stay the same, after any action that is not one of the four aforementioned actions.
Finally, at line $12$, we state that threads are only allowed to be either present in exclusive or are present in shared.

The formula for property~\ref{item:prop_mutex_exclusive} is shown in Table~\ref{prop:busy_forbidden_prop1} and states that when a thread enters the exclusive section, no other thread may enter that section till it leaves the section.
The formulas for properties~\ref{item:prop_event_shared} and~\ref{item:prop_event_crit} are presented in Tables~\ref{prop:busy_forbidden_prop3exclusive} and~\ref{prop:busy_forbidden_prop3shared} and use data parameters to count the number of threads in the exclusive section or in any section respectively.
Note that these are two subformulas with identical structure for shared and exclusive sections respectively.
Finally, property~\ref{item:prop_enter} presented in Table~\ref{prop:busy_forbidden_prop5} uses boolean parameters to keep track of whether any thread is in the shared or exclusive sections, respectively.
This is more efficient than keeping track of the exact amount of threads.

The properties were verified for up to 7 threads.
The specification model has about three million states and the implementation model about 11 billion states.

\begin{table}[H]
\begin{tabular}{|p{\textwidth}|} \hline
$\textit{EnterShared}(p : P) =$\\
\tab{}{1}$\texttt{enter\_shared\_call}(p)~\sequential$\\
\tab{}{1}$\textit{TryBothFlags}(p) ~\sequential$\\
\tab{}{1}$\texttt{enter\_shared\_return}(p)$\\
\\
$\textit{TryBothFlags}(p : P) =$\\
\tab{}{1}$\textit{store}_p(\textit{Busy}(p), \true, p) \sequential ($\\
\tab{}{2}$\textit{load}_p(\textit{Forbidden}(p),~ \true,~ p)~\sequential$\\
\tab{}{2}$\textit{store}_p(\textit{Busy}(p), \false, p) \sequential \textit{improbable} \sequential \textit{TryBothFlags}(p)$\\
\tab{+}{2}$\textit{load}_p(\textit{Forbidden}(p), \false, p)~)$\\ \hline
\end{tabular}
\caption{mCRL2 specification for the implementation of \texttt{enter\_shared}.}
\label{spec:enter_shared}
\end{table}

\begin{table}[H]
\begin{tabular}{|p{\textwidth}|} \hline
$\textit{LeaveShared}(p : P) =$\\
\tab{}{1}$\texttt{leave\_shared\_call}(p)~\sequential$\\
\tab{}{1}$\textit{store}_p(\textit{Busy}(p), \false, p)~\sequential$\\
\tab{}{1}$\texttt{leave\_shared\_return}(p)$ \\ \hline
\end{tabular}
\caption{mCRL2 specification for the implementation of \texttt{leave\_shared}.}
\label{spec:leave_shared}
\end{table}
\vspace{-1cm}
\begin{table}[H]
\begin{tabular}{|p{\textwidth}|} \hline
$\textit{EnterExclusive}(p : P) =$\\
\tab{}{1}$\texttt{enter\_exclusive\_call}(p)$\\
\tab{}{1}$\textit{lock}_p(p)~\sequential$\\
\tab{}{1}$\textit{SetAllForbiddenFlags}(p,~ \emptyset)~\sequential$\\
\tab{}{1}$\texttt{enter\_exclusive\_return}(p)$\\
\\
$\textit{SetAllForbiddenFlags}(p : P,~ \textit{forbidden} : Set(P)) =$\\
\tab{}{1}$(\forall_{p'{:}P}. p \in \textit{forbidden})$\\
\tab{$\rightarrow$}{2}$internal$\\
\tab{$\diamond$}{2}$\sum_{p'{:}P}. \textit{store}_p(\textit{Forbidden}(p'), \true, p) \sequential ($\\
\tab{}{3}$\textit{load}_p(\textit{Busy}(p'), \false, p)~\sequential$\\
\tab{}{3}$\textit{SetAllForbiddenFlags}(p,~ \textit{forbidden} \cup \{p'\})$\\
\tab{+}{3}$l\textit{oad}_p(\textit{Busy}(p'), \true, p)~\sequential$\\
\tab{}{3}$\textit{store}_p(\textit{Forbidden}(p'), \false, p) \sequential \textit{improbable}~\sequential$\\
\tab{}{3}$\textit{SetAllForbiddenFlags}(p,~ \textit{forbidden} \setminus \{ p'\})$\\
\tab{+}{3}$\textit{store}_p(\textit{Forbidden}(p'), \false, p) \sequential \textit{improbable}~\sequential$\\
\tab{}{3}$\textit{SetAllForbiddenFlags}(p,~ \textit{forbidden} \setminus \{p'\})~)$\\ \hline
\end{tabular}
\caption{mCRL2 specification for the \texttt{enter\_exclusive} function in Table~\ref{table:busy_forbidden}.}
\label{spec:enter_exclusive}
\end{table}

\begin{table}[H]
\begin{tabular}{|p{\textwidth}|} \hline
$\textit{LeaveExclusive}(p : P) =$\\
\tab{}{1}$\texttt{leave\_exclusive\_call}(p)~\sequential$\\
\tab{}{1}$\textit{AllowAllThreads}(p,\emptyset)~\sequential$\\
\tab{}{1}$\textit{unlock}_p(p)~\sequential$\\
\tab{}{1}$\texttt{leave\_exclusive\_return}(p)$\\
\\
$\textit{AllowAllThreads}(p : P,~\textit{allowed} : \textit{Set}(P)) =$\\
\tab{}{1}$(\forall_{q{:}P}. q \in \textit{allowed})$\\
\tab{$\rightarrow$}{2}\textit{internal}\\
\tab{$\diamond$}{2}$\sum_{p'{:}P}.($\\
\tab{}{2}$\textit{store}_p(\textit{Forbidden}(p'), \false, p)~\sequential$\\
\tab{}{2}$\textit{AllowAllThreads}(p,~\textit{allowed}\cup\{p'\})$\\
\tab{+}{2}$\textit{store}_p(\textit{Forbidden}(p'), \true, p) \sequential \textit{improbable}~$\\
\tab{}{2}$\textit{AllowAllThreads}(p,~\textit{allowed}\setminus\{p'\}~))$\\ \hline
\end{tabular}
\caption{mCRL2 specification for the \texttt{leave\_exclusive} function in Table~\ref{table:busy_forbidden}.}
\label{spec:leave_exclusive}
\end{table}

\begin{table}[H]
\begin{tabular}{|p{\textwidth}|} \hline
$\textit{Flags}( \textit{flags} : F \rightarrow \boolsort) =$\\
$\sum_{f{:}F,p{:}P}. ($\\
\tab{}{1}$\sum_{b{:}\boolsort}. \textit{store}_f(f, b, p) \sequential \textit{Flag}(\textit{flags}[f\mapsto b])$\\
\tab{+}{1}$\textit{load}_f(f, \textit{flags}(f), p) \sequential \textit{Flag}(\textit{flags})$\\
\tab{}{1}$)$\\
\\
$\textit{Mutex}( \textit{locked}: \boolsort) =$\\
$\sum_{p{:}P}. ($\\
\tab{}{1}$\textit{locked}$\\
\tab{$\rightarrow$}{2}$\textit{lock}_m(p) \sequential \textit{Mutex}(\true)$\\
\tab{$\diamond$}{2}$\textit{unlock}_m(p) \sequential \textit{Mutex}(\false)~)$\\ \hline
\end{tabular}
\caption{mCRL2 specifications for the atomic flags and the mutex.}
\label{spec:busy_forbidden_comm}
\end{table}

\begin{table}[H]
\begin{tabular}{|p{\textwidth}|} \hline
$\textit{Thread}(p : P)=$\\
\tab{}{1}$\textit{EnterShared}(p)~\sequential$\\
\tab{}{1}$\textit{LeaveShared}(p)~\sequential$\\
\tab{}{1}$\textit{Thread}(p)$\\
\tab{+}{1}$\textit{EnterExclusive}(p)~\sequential$\\
\tab{}{1}$\textit{LeaveExclusive}(p)~\sequential$\\
\tab{}{1}$\textit{Thread}(p)$\\ \hline
\end{tabular}
\caption{mCRl2 specification for a thread \textit{p} interacting with the protocol.}
\label{spec:busy_forbidden_thread}
\end{table}

\begin{table}[H]
\begin{tabular}{|p{\textwidth}|} \hline
	$\textbf{allow}(\{$\\
	\tab{}{1}$\textit{store}, \textit{load},$\\
	\tab{}{1}$\textit{lock}, \textit{unlock},$\\
	\tab{}{1}$\textit{internal, improbable,}$\\
	\tab{}{1}$\texttt{enter\_shared\_call}, \texttt{enter\_shared\_return},$\\
	\tab{}{1}$\texttt{leave\_shared\_call}, \texttt{leave\_shared\_return},$\\
	\tab{}{1}$\texttt{enter\_exclusive\_call}, \texttt{enter\_exclusive\_return},$\\
	\tab{}{1}$\texttt{leave\_exclusive\_call}, \texttt{leave\_exclusive\_return}$\\
	\tab{}{1}$\}, \textbf{comm}(\{$\\
	\tab{}{2}$\textit{store}_f | \textit{store}_p \rightarrow \textit{store},$\\
	\tab{}{2}$\textit{load}_f | \textit{load}_p \rightarrow \textit{load},$\\
	\tab{}{2}$\textit{lock}_m | \textit{lock}_p \rightarrow \textit{lock},$\\
	\tab{}{2}$\textit{unlock}_m | \textit{unlock}_p \rightarrow \textit{unlock}$\\
	\tab{}{2}$\}, $\\
    \tab{}{2}$\textit{Thread}(p_1)~||$\\
    \tab{}{3}$\vdots$\\
    \tab{}{2}$\textit{Thread}(p_{|P|})~||$\\
    \tab{}{2}$\textit{Flags}(\lambda f{:}F.\false)~||$\\
    \tab{}{2}$\textit{Mutex}(\false)~)~)$\\ \hline
\end{tabular}
\caption{mCRL2 specification for the busy-forbidden protocol.}
\label{spec:busy_forbidden_init}
\end{table}
\vspace{-1.5cm}
\begin{table}[ht]
\begin{tabular}{|p{\textwidth}|} \hline
$[\true^*]$\\
$[\exists_{p \in P} : \texttt{enter\_exclusive\_return}(p)]$\\
$[\overline{\exists_{p \in P} : \texttt{leave\_exclusive\_call}(p)}^{\,*}]$\\
$[\exists_{p \in P} : \texttt{enter\_exclusive\_return}(p)]$\\
$\false$ \\ \hline
\end{tabular}
\caption{Modal formula for property~\ref{item:prop_mutex_exclusive}: ``There should never be  more than one thread present in the exclusive section''.}
\label{prop:busy_forbidden_prop1}
\end{table}
\vspace{-0.5cm}
\begin{table}[hb]
\begin{tabular}{|p{\textwidth}|} \hline
$\nu X(n_{shared} : \natsort = 0,~n_{exclusive} : \natsort = 0).$\\
\tab{}{1}$(\forall_{p{:}P}. [\texttt{enter\_shared\_return}(p)]X(n_{shared}+1, n_{exclusive})~)$\\
\tab{$\wedge$}{1}$(\forall _{p{:}P}.[\texttt{enter\_exclusive\_return}(p)]X(n_{shared}, n_{exclusive}+1)~)$\\
\tab{$\wedge$}{1}$(\forall_{p{:}P}.[\texttt{leave\_shared\_call}(p)]X(n_{shared}-1, n_{exclusive})~)$\\
\tab{$\wedge$}{1}$(\forall _{p{:}P}.[\texttt{leave\_exclusive\_call}(p)]X(n_{shared}, n_{exclusive}-1)~)$\\
\tab{$\wedge$}{1}\tab{$[$}{0}$\overline{(\exists _{p{:}P}. \texttt{enter\_shared\_return}(p)~)}$\\
\tab{$\cap$}{2}$\overline{(\exists _{p{:}P}. \texttt{enter\_exclusive\_return}(p)~)}$\\
\tab{$\cap$}{2}$\overline{(\exists _{p{:}P}. \texttt{leave\_shared\_call}(p)~)}$\\
\tab{$\cap$}{2}$\overline{(\exists _{p{:}P}. \texttt{leave\_exclusive\_call}(p)~)}$\\
\tab{}{1}$]X(n_{shared}, n_{exclusive})$\\
\tab{$\wedge$}{1}$\neg (n_{exclusive} > 0 \land n_{shared} > 0)$\\ \hline
\end{tabular}
\caption{The modal formula for property~\ref{item:prop_crtical_mutex_w_shared}: ``There should never be a thread present in the exclusive section while one or more threads are present in the shared section''.}
\label{property:busy_forbidden_p2}
\end{table}

\begin{table}[H]
\begin{tabular}{|p{\textwidth}|} \hline
$\nu X(n_{exclusive} : \natsort = 0).$\\
\tab{}{1}$[\exists _{p{:}P}.~\texttt{enter\_exclusive\_call}(p)]X(n_{exclusive}+1)$\\
\tab{$\wedge$}{1}$[\exists _{p{:}P}.~\texttt{leave\_exclusive\_return}(p)]X(n_{exclusive}-1)$\\
\tab{$\wedge$}{1}\tab{$[$}{0}$\overline{(\exists _{p{:}P}.~\texttt{enter\_exclusive\_call}(p))}$\\
\tab{$\cap$}{2}$\overline{(\exists _{p{:}P}.~\texttt{leave\_exclusive\_return}(p))}$\\
\tab{}{1}$]~X(n_{exclusive})$\\
\tab{$\wedge$}{1}$\forall _{p{:}P}. [\texttt{enter\_shared\_call}(p)]$\\
\tab{}{1}$~\nu Y(n_{exclusive}' : \natsort = n_{exclusive}). \mu Z(n_{exclusive}'' : \natsort = n_{exclusive}').~($\\
\tab{}{2}\tab{$[$}{0}$\overline{\texttt{enter\_shared\_return}(p)}$\\
\tab{$\cap$}{3}$\overline{(\exists _{p'{:}P}.~\texttt{enter\_shared\_call}(p'))}$\\
\tab{$\cap$}{3}$\overline{(\exists _{p'{:}P}.~\texttt{enter\_exclusive\_call}(p'))}$\\
\tab{$\cap$}{3}$\overline{(\exists _{p'{:}P}.~\texttt{leave\_exclusive\_return}(p'))}$\\
\tab{$\cap$}{3}$\overline{\textit{improbable}}$\\
\tab{$]~($}{3}$((n_{exclusive}'' \approx 0)  \implies Z(n_{exclusive}''))$\\
\tab{$\wedge$}{3}$((n_{exclusive}'' > 0)  \implies Y(n_{exclusive}''))~)$\\
\tab{$\wedge$}{2}$[\exists _{p'{:}P}.~\texttt{enter\_shared\_call}(p')]Y(n_{exclusive}'')$\\
\tab{$\wedge$}{2}$[\exists _{p'{:}P}.~\texttt{enter\_exclusive\_call}(p')]Y(n_{exclusive}''+1)$\\
\tab{$\wedge$}{2}$[\exists _{p'{:}P}.~\texttt{leave\_exclusive\_return}(p')]Y(n_{exclusive}''-1)$\\
\tab{$\wedge$}{2}$[\textit{improbable}]Y(n_{exclusive}'')$\\
\tab{$\wedge$}{2}$\langle \true^* \sequential \texttt{enter\_shared\_return}(p)\rangle\true~)$\\ \hline
\end{tabular}

\caption{Modal formula for property~\ref{item:prop_event_shared}: ``When a thread requests to enter the shared section, it will be granted access within a bounded number of steps, unless there is another thread in the exclusive section''.}
\label{prop:busy_forbidden_prop3exclusive}
\end{table}
\vspace{-3cm}
\begin{table}[ht]
\begin{tabular}{|p{\textwidth}|} \hline
$[\true^*]~\forall _{p{:}P}.($\\
\tab{}{1}$[\texttt{leave\_shared\_call}(p)]\nu X.\mu Y. ($\\
\tab{$[$}{3}$\overline{\texttt{leave\_shared\_return}(p)}$\\
\tab{$\cap$}{3}$\overline{(\exists _{p'{:}P}.~\texttt{enter\_exclusive\_call}(p'))}$\\
\tab{$\cap$}{3}$\overline{(\exists _{p'{:}P}.~\texttt{enter\_shared\_call}(p'))}$\\
\tab{$\cap$}{3}$\overline{\textit{improbable}}~]~Y$\\
\tab{$\wedge$}{2}\tab{$[$}{0}$(\exists _{p'{:}P}.~\texttt{enter\_exclusive\_call}(p'))$\\
\tab{$\cup$}{3}$(\exists _{p'{:}P}.~\texttt{enter\_shared\_call}(p'))$\\
\tab{$\cup$}{3}$(\textit{improbable})~]~X$\\
\tab{$\wedge$}{2}$\langle \true^* \sequential \texttt{leave\_shared\_return}(p)\rangle\true~)$\\
\tab{$\wedge$}{1}$[\texttt{leave\_exclusive\_call}(p)]\nu X.\mu Y. ($\\
\tab{$[$}{3}$\overline{\texttt{leave\_exclusive\_return}(p)}$\\
\tab{$\cap$}{3}$\overline{(\exists _{p'{:}P}.~\texttt{enter\_exclusive\_call}(p'))}$\\
\tab{$\cap$}{3}$\overline{(\exists _{p'{:}P}.~\texttt{enter\_shared\_call}(p'))}$\\
\tab{$\cap$}{3}$\overline{\textit{improbable}}~]~Y$\\
\tab{$\wedge$}{2}\tab{$[$}{0}$(\exists _{p'{:}P}.~\texttt{enter\_exclusive\_call}(p'))$\\
\tab{$\cup$}{3}$(\exists _{p'{:}P}.~\texttt{enter\_shared\_call}(p'))$\\
\tab{$\cup$}{3}$(\textit{improbable})~]~X$\\
\tab{$\wedge$}{2}$\langle \true^* \sequential \texttt{leave\_exclusive\_return}(p)\rangle\true~)~)$\\ \hline
\end{tabular}
\caption{Modal formula for property~\ref{item:prop_leave}: ``When a thread requests to leave the exclusive/shared section, it will leave it within a bounded number of steps''.}
\label{prop:busy_forbidden_prop4}
\end{table}

\begin{table}[H]
\begin{tabular}{|p{\textwidth}|} \hline
$\nu X(n_{blocking} : \natsort = 0).$\\
\tab{}{1}$[\exists _{p{:}P}.~\texttt{enter\_exclusive\_call}(p)]X(n_{blocking}+1)$\\
\tab{$\wedge$}{1}$[\exists _{p{:}P}.~\texttt{enter\_shared\_call}(p)]X(n_{blocking}+1)$\\
\tab{$\wedge$}{1}$[\exists _{p{:}P}.~\texttt{leave\_shared\_return}(p)]X(n_{blocking}-1)$\\
\tab{$\wedge$}{1}$[\exists _{p{:}P}.~\texttt{leave\_exclusive\_return}(p)]X(n_{blocking}-1)$\\
\tab{$\wedge$}{1}\tab{$[$}{0}$\overline{(\exists _{p{:}P}.~\texttt{enter\_exclusive\_call}(p))}$\\
\tab{$\cap$}{2}$\overline{(\exists _{p{:}P}.~\texttt{leave\_exclusive\_return}(p))}$\\
\tab{$\cap$}{2}$\overline{(\exists _{p{:}P}.~\texttt{enter\_shared\_call}(p))}$\\
\tab{$\cap$}{2}$\overline{(\exists _{p{:}P}.~\texttt{leave\_shared\_return}(p))}$\\
\tab{}{1}$]~X(n_{exclusive})$\\
\tab{$\wedge$}{1}$\forall _{p{:}P}. [\texttt{enter\_exclusive\_call}(p)]$\\
\tab{}{1}$~\nu Y(n_{blocking}' : \natsort = n_{blocking}). \mu Z(n_{blocking}'' : \natsort = n_{blocking}').~($\\
\tab{$[$}{3}$\overline{\texttt{enter\_exclusive\_return}(p)}$\\
\tab{$\cap$}{3}$\overline{(\exists _{p'{:}P}.~\texttt{enter\_shared\_call}(p))}$\\
\tab{$\cap$}{3}$\overline{(\exists _{p'{:}P}.~\texttt{leave\_shared\_return}(p))}$\\
\tab{$\cap$}{3}$\overline{(\exists _{p'{:}P}.~\texttt{enter\_exclusive\_call}(p'))}$\\
\tab{$\cap$}{3}$\overline{(\exists _{p'{:}P}.~\texttt{leave\_exclusive\_return}(p'))}$\\
\tab{$\cap$}{3}$\overline{\textit{improbable}}$\\
\tab{$]~($}{3}$((n_{blocking}'' \approx 0)  \implies Z(n_{blocking}''))$\\
\tab{$\wedge$}{3}$((n_{blocking}'' > 0)  \implies Y(n_{blocking}''))~)$\\
\tab{$\wedge$}{2}$[\exists _{p'{:}P}.~\texttt{enter\_shared\_call}(p')]Y(n_{blocking}''+1)$\\
\tab{$\wedge$}{2}$[\exists _{p'{:}P}.~\texttt{leave\_shared\_return}(p')]Y(n_{blocking}''-1)$\\
\tab{$\wedge$}{2}$[\exists _{p'{:}P}.~\texttt{enter\_exclusive\_call}(p')]Y(n_{blocking}''+1)$\\
\tab{$\wedge$}{2}$[\exists _{p'{:}P}.~(p'\not\approx p)~\wedge~\texttt{enter\_exclusive\_return}(p')]Y(n_{blocking}''-1)$\\
\tab{$\wedge$}{2}$[\textit{improbable}]Y(n_{exclusive}'')$\\
\tab{$\wedge$}{2}$\langle \true^* \sequential \texttt{enter\_exclusive\_return}(p)\rangle\true~)$\\ \hline
\end{tabular}
\caption{Modal formula for property \ref{item:prop_event_crit}: ``When a thread requests to enter the exclusive section, it will be granted access within a bounded number of steps, unless there is another thread in the shared or
in the exclusive section''.}
\label{prop:busy_forbidden_prop3shared}
\end{table}
\vspace{-3cm}

\begin{table}[ht]
\begin{tabular}{|p{\textwidth}|} \hline
$\forall _{p{:}P}.\nu X(b_{shared} : \boolsort = \false,~b_{exclusive} : \boolsort = \false).$\\
\tab{}{1}$[\texttt{enter\_shared\_call}(p)]X(\true,b_{exclusive})$\\
\tab{$\wedge$}{1}$[\texttt{leave\_shared\_return}(p)]X(\false,b_{exclusive})$\\
\tab{$\wedge$}{1}$[\texttt{enter\_exclusive\_call}(p)]X(b_{shared},\true)$\\
\tab{$\wedge$}{1}$[\texttt{leave\_exclusive\_return}(p)]X(b_{shared},\false)$\\
\tab{$\wedge$}{1}\tab{$[$}{0}$\overline{\texttt{enter\_shared\_call}(p)}$\\
\tab{$\cap$}{2}$\overline{\texttt{leave\_shared\_return}(p)}$\\
\tab{$\cap$}{2}$\overline{\texttt{enter\_exclusive\_call}(p)}$\\
\tab{$\cap$}{2}$\overline{\texttt{leave\_exclusive\_return}(p)}$\\
\tab{}{1}$]~X(n_{shared},n_{exclusive})$\\
\tab{$\wedge$}{1}$((\neg n_{shared} \wedge \neg n_{exclusive})  \implies ($\\
\tab{}{2}$\langle \texttt{enter\_exclusive\_call(p)}\rangle\true$\\
\tab{$\wedge$}{2}$\langle \texttt{enter\_shared\_call(p)}\rangle\true)~)$\\ \hline
\end{tabular}
\caption{Modal formula for property~\ref{item:prop_enter}: ``A thread not in the exclusive or shared section can instantly start to enter the exclusive or shared section''.}
\label{prop:busy_forbidden_prop5}
\end{table}

\section{mCRL2 Specifications for the Term Library} \label{app:aterm-impl}
In this section we give the formal mCRL2 specifications of the implementation of the term library that is used 
to perform model checking.
Creating a term is specified in Table~\ref{spec:atermpp_create} and destroying a term in Table~\ref{spec:atermpp_destroy}.
In this model, the set $P$ corresponds to the set containing all threads, $T$ to the set containing all terms and $A$ to the set containing all memory addresses. 
The set $A_\bot = A \cup \{\bot\}$ with $\bot \not\in A$ contains the extra element $\bot$ meaning no address or a NULL pointer.
To ensure finiteness and reduce the complexity of the model, the set $T$ only contains a finite amount of constants, \ie, terms of arity zero.

\begin{table}[h]
\begin{tabular}{|p{\textwidth}|} \hline
$\textit{Create}(p: P,~t:T,~\textit{lm} : T \rightarrow A_\bot) =$\\
\tab{}{1}$\texttt{create\_call}(p,~t)~\sequential$\\
\tab{}{1}$\textit{EnterShared}(p)~\sequential$\\
\tab{}{1}$\textit{Create}_2(p,~t,~\textit{lm})$\\
\\
$\textit{Create}_2(p: P,~t:T,~\textit{lm} : T \rightarrow A_\bot) =$\\
$\sum _{a{:}A_\bot}. ($\\
\tab{}{1}$\textit{contains}_p(t,a,p)~\sequential$\\
\tab{}{1}$(a \approx \bot)$\\
\tab{$\rightarrow$}{2}$\sum_{a'{:}A}.~($\\
\tab{}{3}$\textit{construct\_term}_p(t~,a'~,p) \sequential ($\\
\tab{}{4}$\textit{insert}_p(t~,a'~,\true~,p)~\sequential$\\
\tab{}{4}$\textit{Create}_3(p,~t,~\textit{lm},~a')$\\
\tab{+}{4}$\textit{insert}_p(t,~a',~\false,~p)~\sequential$\\
\tab{}{4}$\textit{destruct\_term}_p(t,~a',~p)~\sequential$\\
\tab{}{4}$\textit{Create}_2(p,~t,~\textit{lm})~)~)$\\
\tab{$\diamond$}{2}$\textit{Create}_3(p,~t,~\textit{lm},~a)~)$\\
\\
$\textit{Create}_3(p:P,t:T,~\textit{lm} : T \rightarrow A_\bot,~a:A) =$\\
\tab{}{1}$\textit{protect}_p(t,~a,~p)~\sequential$\\
\tab{}{1}$\textit{LeaveShared}(p)~\sequential$\\
\tab{}{1}$\texttt{create\_return}(p,~t,~a)~\sequential$\\
\tab{}{1}$\textit{Thread}(p,~\textit{lm}[t \mapsto a])$\\ \hline
\end{tabular}
\caption{mCRL2 specification for the $\textsf{create}$ function shown in Table~\ref{table:par_aterm}.}
\label{spec:atermpp_create}
\end{table}

First of all, we introduce processes $\mathit{EnterShared}$,  $\mathit{LeaveShared}$,  $\mathit{EnterExclusive}$ and  $\mathit{LeaveExclusive}$ to interact with the busy-forbidden specification $\mathit{BF}$ specified in Table~\ref{spec:busy_forbidden}.
To distinguish between the term library and the protocol all actions such as $\texttt{enter\_shared\_call}$ are split into action $\texttt{enter\_shared\_call}_\textit{bf}$ for the protocol and $\texttt{enter\_shared\_call}_\textit{p}$ for the term library.
Finally, we have the process $\mathit{MainMemory}$ to model the main memory by keeping track of \emph{used} memory addresses, the process $\mathit{HashTable}$ to model a hash table as an associative array, and process 
$\mathit{ReferenceCounter}$ to track a reference counter for every address (or term).
Destroying a term is specified in Table~\ref{spec:atermpp_destroy}, which uses the same other processes as the creation function.
Again, there are two separate processes to model the behaviour of the while loop.

The specification in Table~\ref{spec:atermpp} models the behaviour of each thread. 
Each thread repeatedly tries to either creates a term it does not yet know, or it destroys a known term. 
Finally, Table~\ref{spec:atermpp_init} shows the complete specification including the communication between various processes used to model the thread-safe term library. 

\begin{table}[H]
\begin{tabular}{|p{\textwidth}|} \hline
$\textit{Destroy}(p:P,~t:T,~\textit{lm} : T \rightarrow A_\bot) =$\\
\tab{}{1}$\texttt{destroy\_call}(p,~t)~\sequential$\\
\tab{}{1}$\textit{unprotect}_p(t,~\textit{lm}(t),~p) \sequential ($\\
\tab{}{2}$skip$\\
\tab{+}{2}$skip \sequential GC(p)~)~\sequential$\\
\tab{}{1}$\texttt{destroy\_return}(p)~\sequential$\\
\tab{}{1}$Thread(p,~\textit{lm}[t \mapsto \bot])$\\
\\
$\textit{GC}(p:P) =$\\
\tab{}{1}$\textit{EnterExclusive}(p)~\sequential$\\
\tab{}{1}$\textit{GC}_2(p,~\emptyset)$\\
\\
$\textit{GC}_2(p:P,~\textit{checked} : \textit{FSet}(T)) =$\\
\tab{}{1}$(\forall _{t{:}T}. t \in \textit{checked})$\\
\tab{$\rightarrow$}{2}$\textit{LeaveExclusive}(p)$\\
\tab{$\diamond$}{2}$\sum _{t{:}T}. (t \not\in \textit{checked}) \rightarrow ($\\
\tab{}{3}$\textit{contains}_p(t,~\bot,~p)~\sequential$\\
\tab{}{3}$\textit{GC}_2(p,~\textit{checked} \cup \{t\}$\\
\tab{+}{3}$\sum _{a{:}A}. \textit{contains}_p(t,~a,~p) \sequential ($\\
\tab{}{4}$\textit{protected}_p(a,~\true,~p)~\sequential$\\
\tab{}{4}$\textit{GC}_2(p,~\textit{checked} \cup \{t\})$\\
\tab{+}{4}$\textit{protected}_p(a,~\false,~p)~\sequential$\\
\tab{}{4}$\textit{destruct\_term}_p(t,~a,~p)~\sequential$\\
\tab{}{4}$\textit{delete}_p(t,~p)~\sequential$\\
\tab{}{4}$\textit{GC}_2(p,~\textit{checked} \cup \{t\})~)~)$\\ \hline
\end{tabular}
\caption{mCRL2 specification for the $\textsf{destroy}$ function shown in Table~\ref{table:par_aterm}.}
\label{spec:atermpp_destroy}
\end{table}

To verify the model of the thread-safe term library we again specify a number of modal formulas for the properties described in Section~\ref{behaviouralproperties}.
The modal formula for property~\ref{item:prop_no_spontaneous_destroy} specified in Table~\ref{prop:atermpp_prop1} uses 
a mapping $a$ from addresses to terms and the finite set $\textit{owners}$ containing all threads that own/protect 
term $t$ as data parameters.
If at any point in time a $\texttt{create(t)}$ returns a different address than the current address, then the term must not be owned by any thread.
The modal formula in Table~\ref{prop:atermpp_prop2} for property~\ref{item:prop_injective} uses the same constructs to check whether terms on the same address are also equivalent.

The formula for property~\ref{item:prop_can_start} shown in Table~\ref{prop:atermpp_prop3} uses a boolean parameter $\textit{busy}$ to keep track of whether the thread $p$ is creating (or destroying) a term.
Furthermore, the parameter $\textit{known}$ is a finite set containing all terms that thread $p$ knows.
If at any point in time $\textit{busy}$ is false, then the process must be able to start destroying any term in $\textit{known}$ and start creating any term not currently in $\textit{known}$.
Finally, for property~\ref{item::finish_eventually} the formula shown in Table~\ref{prop:atermpp_prop4} uses again the construction which (under fairness) indicates that term creation and destruction will finish within a finite number of steps.
Note that the subformulas for creation and destruction have an identical structure.

\begin{table}[H]
\begin{tabular}{|p{\textwidth}|} \hline
$\textit{MainMemory}(\textit{used} : \textit{FSet}(A)) =$\\
$\sum_{p{:}P,t{:}T,a{:}A}.((a \not\in \textit{used})$\\
\tab{$\rightarrow$}{2}$\textit{construct\_term}_{mm}(t,a,p) \sequential \textit{MainMemory}(\textit{used} \cup \{a\})$\\
\tab{$\diamond$}{2}$\textit{destruct\_term}_{mm}(t,a,p) \sequential \textit{MainMemory}(\textit{used} \setminus \{a\})~)$\\
\\
$\textit{HashTable}(m : T \rightarrow A_\bot) =$\\
$\sum_{t{:}T,p{:}P}. ($\\
\tab{}{1}$\textit{contains}_{ht}(t, m(e), p) \sequential \textit{HashTable}(m)$\\
\tab{+}{1}$\sum _{a{:}A}. (m(e) \approx \bot)$\\
\tab{$\rightarrow$}{2}$\textit{insert}_{ht}(t,a,\true, p) \sequential \textit{HashTable}(m[e \mapsto a)$\\
\tab{$\diamond$}{2}$\textit{insert}_{ht}(t,a,\false,p) \sequential \textit{HashTable}(m)$\\
\tab{+}{1}$\textit{delete}_{ht}(t, p) \sequential \textit{HashTable}(m[e \mapsto \bot])~)$\\
\\
$\textit{ReferenceCounter}(counter : A \rightarrow \natsort) =$\\
\tab{}{1}$\sum_{t{:}T,p{:}P}.~\textit{protect}_{rc}(t,a,p)~\sequential$\\
\tab{}{1}$\textit{ReferenceCounter}(counter[a\mapsto counter(a)+1]$\\
\tab{+}{1}$\sum_{t{:}T,p{:}P}.~\textit{unprotect}_{rc}(t,a,p)~\sequential$\\
\tab{}{1}$\textit{ReferenceCounter}(counter[a\mapsto counter(a)-1]$\\
\tab{+}{1}$\sum_{t{:}T,p{:}P}.~\textit{protected}_{rc}(t,a,counter(a)\not\approx 0,p)~\sequential$\\
\tab{}{1}$\textit{ReferenceCounter}(counter)$\\ \hline
\end{tabular}
\caption{mCRL2 specifications of the main memory, hash table and reference counters used in the term library specification.}
\label{spec:atermpp_comm}
\end{table}
\vspace{-3cm}
\begin{table}[ht]
\begin{tabular}{|p{\textwidth}|} \hline
$Thread(p:P,~\textit{lm} : T \rightarrow A_\bot) =$\\
\tab{}{1}$(\sum _{t{:}T}. (\textit{lm}(t) \approx \bot) \rightarrow \textit{Create}(p,t,\textit{lm}))$\\
\tab{+}{1}$(\sum _{t{:}T}. (\textit{lm}(t) \not\approx \bot) \rightarrow \textit{Destroy}(p,t,\textit{lm}))$\\ \hline
\end{tabular}
\caption{mCRL2 specification of a thread $p$ interacting with the term library.}
\label{spec:atermpp}\setcounter{table}{\value{table}+1}
\end{table}

\begin{table}[H]
\begin{tabular}{|p{\textwidth}|} \hline
\textbf{allow}$(\{$\\
\tab{}{1}$\textit{construct\_term}, \textit{destruct\_term},$\\
\tab{}{1}$\textit{contains}, \textit{insert}, \textit{delete},$\\
\tab{}{1}$\textit{protect}, \textit{unprotect}, \textit{protected},$\\
\tab{}{1}$\textit{skip, improbable},$\\
\tab{}{1}$\textit{enter\_shared\_call}, \textit{enter\_shared\_return},$\\
\tab{}{1}$\textit{leave\_shared\_call}, \textit{leave\_shared\_return},$\\
\tab{}{1}$\textit{enter\_exclusive\_call}, \textit{enter\_exclusive\_return},$\\
\tab{}{1}$\textit{leave\_exclusive\_call},\textit{leave\_exclusive\_return},$\\
\tab{}{1}$\texttt{create\_call}, \texttt{create\_return},$\\
\tab{}{1}$\texttt{destroy\_call}, \texttt{destroy\_return}$\\
\tab{}{1}$\}, \textbf{comm}(\{$\\
\tab{}{2}$\textit{construct\_term}_{mm} | \textit{construct\_term}_p \rightarrow \textit{construct\_term},$\\
\tab{}{2}$\textit{destruct\_term}_{mm} | \textit{destruct\_term}_p \rightarrow \textit{destruct\_term}, $\\
\tab{}{2}$\textit{contains}_{ht} | \textit{contains}_p \rightarrow \textit{contains},$\\
\tab{}{2}$\textit{insert}_{ht} | \textit{insert}_p \rightarrow \textit{insert},$\\
\tab{}{2}$\textit{delete}_{ht} | \textit{delete}_p \rightarrow \textit{delete},$\\
\tab{}{2}$\textit{protect}_{rc} | \textit{protect}_p \rightarrow \textit{protect},$\\
\tab{}{2}$\textit{unprotect}_{rc} | \textit{unprotect}_p \rightarrow \textit{unprotect},$\\
\tab{}{2}$\textit{protected}_{rc} | \textit{protected}_p \rightarrow \textit{protected},$\\
\tab{}{2}$\textit{enter\_shared\_call}_{\textit{bf}} | \textit{enter\_shared\_call}_p \rightarrow \textit{enter\_shared\_call},$\\
\tab{}{2}$\textit{enter\_shared\_return}_{\textit{bf}} | \textit{enter\_shared\_return}_p \rightarrow \textit{enter\_shared\_return},$\\
\tab{}{2}$\textit{leave\_shared\_call}_{\textit{bf}} | \textit{leave\_shared\_call}_p \rightarrow \textit{leave\_shared\_call},$\\
\tab{}{2}$\textit{leave\_shared\_return}_{\textit{bf}} | \textit{leave\_shared\_return}_p \rightarrow \textit{leave\_shared\_return},$\\
\tab{}{2}$\textit{enter\_exclusive\_call}_{\textit{bf}} | \textit{enter\_exclusive\_call}_p \rightarrow \textit{enter\_exclusive\_call},$\\
\tab{}{2}$\textit{enter\_exclusive\_return}_{\textit{bf}} | \textit{enter\_exclusive\_return}_p \rightarrow \textit{enter\_exclusive\_return},$\\
\tab{}{2}$\textit{leave\_exclusive\_call}_{\textit{bf}} | \textit{leave\_exclusive\_call}_p \rightarrow \textit{leave\_exclusive\_call},$\\
\tab{}{2}$\textit{leave\_exclusive\_return}_{\textit{bf}} | \textit{leave\_exclusive\_return}_p \rightarrow \textit{leave\_exclusive\_return}~\},$\\
\tab{}{2}$\textit{Thread}(p_1)~||$\\
\tab{}{3}$\vdots$\\
\tab{}{2}$\textit{Thread}(p_{|P|})~||$\\
\tab{}{2}$\textit{MainMemory}(\emptyset)~||$\\
\tab{}{2}$\textit{HashTable}(\lambda t{:}T.\bot)~||$\\
\tab{}{2}$\textit{ReferenceCounter}(\lambda a{:}A.~0)~||$\\
\tab{}{2}$\textit{BF}(\lambda p{:}P.\textit{Free})~)~)$\\ \hline
\end{tabular}
\caption{mCRL2 specification for the thread-safe term library.}
\label{spec:atermpp_init}
\end{table}
\vspace{-2cm}

\begin{table}[ht]
\begin{tabular}{|p{\textwidth}|} \hline
$\forall _{t{:}T}. \nu X(a:A_\bot = \bot,~\textit{owners} : \textit{FSet}(P) = \emptyset).$\\
\tab{}{1}$(\forall_{p{:}P, a'{:}A}.$\\
\tab{}{2}$[\texttt{create\_return}(p,t,a')]~($\\
\tab{}{3}$X(a', \textit{owners} \cup \{p\})$\\
\tab{$\wedge$}{3}$(a \not\approx a' \implies \textit{owners} \approx \emptyset)~)~)$\\
\tab{$\wedge$}{1}$(\forall p:P.~[\texttt{destroy\_call}(p,t)]X(a,\textit{owners} \setminus \{p\}))$\\
\tab{$\wedge$}{1}\tab{$[$}{0}$\overline{\exists_{p{:}P, a'{:}A}. \texttt{create\_return}(p,t,a')}$\\
\tab{$\cap$}{2}$\overline{\exists_{p{:}P}. \texttt{destroy\_call}(p, t)}$\\
\tab{}{1}$]~X(a,\textit{owners})$ \\ \hline
\end{tabular}
\caption{Formulation of property \ref{item:prop_no_spontaneous_destroy}: ``A term and all its subterms remain in existence at exactly the same address, with unchanged function symbol and arguments, as long as it is not destroyed''.}
\label{prop:atermpp_prop1}
\end{table}

\begin{table}[ht]
\begin{tabular}{|p{\textwidth}|} \hline
$\forall _{a{:}A, t_1{:}T}. \nu X(t:T = t_1,\textit{owners} : \textit{FSet}(P) = \emptyset).$\\
\tab{}{1}$(\forall _{p{:}P, t_2{:}T}.$\\
\tab{}{2}$[\texttt{create\_return}(p,t_2,a)]~($\\
\tab{}{3}$X(t_2, \textit{owners} \cup \{p\})$\\
\tab{$\wedge$}{3}$(t \not\approx t_2 \implies \textit{owners} \approx \emptyset)~)~)$\\
\tab{$\wedge$}{1}$(\forall _{p{:}P}.~[\texttt{destroy\_call}(p,t)]X(t,\textit{owners} \setminus \{p\}))$\\
\tab{$\wedge$}{1}\tab{$[$}{0}$\overline{\exists _{p{:}P, t'{:}T}. \texttt{create\_return}(p,t',a)}$\\
\tab{$\cap$}{2}$\overline{\exists _{p{:}P}. \texttt{destroy\_call}(p,t)}$\\
\tab{}{1}$]~X(t,\textit{owners})$ \\ \hline
\end{tabular}
\caption{Modal formula for property \ref{item:prop_injective}: ``Two stored terms $t_1$ and $t_2$ always have the same non-null address iff they are equal''.}
\label{prop:atermpp_prop2}
\end{table}

\begin{table}[H]
\begin{tabular}{|p{\textwidth}|} \hline
$\forall _{p{:}P}.~\nu X(\textit{busy} : \boolsort = \false, \textit{known} : \textit{FSet}(T) = \emptyset).$\\
\tab{}{1}$(\neg \textit{busy}) \rightarrow ($\\
\tab{}{2}$(\forall _{t{:}T}. (t \not\in \textit{known}) \implies [\tau^*]\langle \tau^*.\texttt{create\_call}(p,t)\rangle\true)$\\
\tab{$\wedge$}{2}$(\forall _{t{:}T}. (t \in known) \implies [ \tau^*] \langle \tau^*.\texttt{destroy\_call}(p,t)\rangle\true)~)$\\
\tab{$\wedge$}{1}\tab{$[$}{0}$(\exists _{t{:}T}. \texttt{create\_call}(p,t))$\\
\tab{$\cup$}{2}$(\exists _{t{:}T}. \texttt{destroy\_call}(p,t))$\\
\tab{}{1}$]~X(\true, \textit{owned})$\\
\tab{$\wedge$}{1}$(\forall _{t{:}T}.~[\exists _{a{:}A}. \texttt{create\_return}(p,t,a)]X(\false, \textit{owned} \cup \{t\}))$\\
\tab{$\wedge$}{1}$(\forall _{t{:}T}.~[\texttt{destroy\_return}(p,t)]X(\false, \textit{owned} \setminus \{t\}))$\\
\tab{$\wedge$}{1}\tab{$[$}{0}$\overline{(\exists _{t{:}T}. \texttt{create\_call}(p,t))}$\\
\tab{$\cap$}{2}$\overline{(\exists _{t{:}T}. \texttt{destroy\_call}(p,t))}$\\
\tab{$\cap$}{2}$\overline{(\exists _{t{:}T,a{:}A}. \texttt{create\_return}(p,t,a))}$\\
\tab{$\cap$}{2}$\overline{(\exists _{t{:}T}. \texttt{destroy\_return}(p,t))}$\\
\tab{}{1}$]~X(\textit{busy},\textit{owned})$ \\ \hline
\end{tabular}
\caption{Modal formula for property \ref{item:prop_can_start}: ``Any thread that is not busy creating or destroying a term, can always initiate the construction of a new term or the destruction of an owned term, \ie, a term that
this thread has exclusive access to''.}
\label{prop:atermpp_prop3}
\end{table}

\begin{table}[H]
\begin{tabular}{|p{\textwidth}|} \hline
$([\true^*]\forall _{p{:}P, t{:}T}.[\texttt{create\_call}(p,t)]~\nu X_c.~\mu Y_c.($\\
\tab{}{1}$\forall _{p'{:}P}. (p \not\approx p') \implies$\\
\tab{}{2}\tab{$[$}{0}$(\exists _{t'{:}T}.~\texttt{create\_call}(p',t'))$\\
\tab{$\cup$}{3}$(\exists _{t'{:}T}.~\texttt{destroy\_call}(p',t'))$\\
\tab{$\cup$}{3}$\textit{improbable}~]~X_c$\\
\tab{$\wedge$}{1}\tab{$[$}{0}$\overline{(\exists _{a{:}A}.~\texttt{create\_return}(p,t,a))}$\\
\tab{$\cap$}{2}$\overline{(\exists _{p'{:}P}.(p \not\approx p') \cap (\exists _{t'{:}T}.~\texttt{create\_call}(p',t')))}$\\
\tab{$\cap$}{2}$\overline{(\exists _{p'{:}P}.(p \not\approx p') \cap (\exists _{t'{:}T}.~\texttt{destroy\_call}(p',t')))}$\\
\tab{$\cap$}{2}$\overline{\textit{improbable}}~]~Y_c$\\
\tab{$\wedge$}{1}$\langle\true^*\sequential\exists _{a{:}A}.~\texttt{create\_return}(p,t,a)\rangle\true~)~)$\\
$\wedge$\\
$([\true^*]\forall_{p{:}P,t{:}T}.[\texttt{destroy\_call}(p,t)]~\nu X_d.~\mu Y_d.($\\
\tab{}{1}$\forall_{p'{:}P}. (p \not\approx p') \implies$\\
\tab{$[$}{3}$(\exists_{t'{:}T}.~\texttt{create\_call}(p',t'))$\\
\tab{$\cup$}{3}$(\exists_{t'{:}T}.~\texttt{destroy\_call}(p',t'))$\\
\tab{$\cup$}{3}$\textit{improbable}~]~X_d$\\
\tab{$\wedge$}{1}\tab{$[$}{0}$\overline{\texttt{destroy\_return}(p,t)}$\\
\tab{$\cap$}{2}$\overline{(\exists_{p'{:}P}.(p \not\approx p') \cap (\exists_{t'{:}T}.~\texttt{create\_call}(p',t')))}$\\
\tab{$\cap$}{2}$\overline{(\exists_{p'{:}P}.(p \not\approx p') \cap (\exists_{t'{:}T}.~\texttt{destroy\_call}(p',t')))}$\\
\tab{$\cap$}{2}$\overline{\textit{improbable}}~]~Y_d$\\
\tab{$\wedge$}{1}$\langle\true^*\sequential\texttt{destroy\_return}(p,t)\rangle\true~)~)$\\ \hline
\end{tabular}
\caption{Modal formula(s) for property \ref{item::finish_eventually}: ``Any thread that started creating a term or destroying a term, will eventually successfully finish this task provided there is enough memory to store one more term than those that are in use. But it is required that other threads behave fairly, in the sense that they will not continually create and destroy terms or stall other threads by busy waiting''.}
\label{prop:atermpp_prop4}
\end{table}

\section{Benchmark data}
The benchmark tests and information shown in Figures \ref{fig:results} and 
\ref{fig::addbenchmarks} are hard to read exactly. Therefore, we repeat the corresponding precise benchmark 
numbers in 
Table \ref{tabel:results_shared_create} up to and including \ref{table:result_statespace_exploration}.
Each wall-clock time is measured in seconds.

The measurements in Table \ref{table:results_creating_shared_intel} came from benchmarking performed on an Intel i7-7700HQ processor.
All other measurements were obtained through benchmarking on an AMD EPYC 7452 32-Core processor.

The benchmark results in Table \ref{tabel:results_shared_create} were obtained by having each thread create 
a term $t_{400~000}$, with $t_0$ being a constant, and $t_{i+1}$ equal to $f(t_i,t_i)$. 
No garbage collection was performed during the benchmark. Note that only one copy of the term
is actually stored in memory. So, most threads wanting to construct some term $f(t_i,t_i)$ detect that
the term already exists, and only need to return its address, without actually creating it. 

The benchmark results in Table \ref{table:results_unique_create} were obtained by having each thread create its
own copy of the term $t_{400~000 / \#\textit{threads}}$, and measuring the wall-clock time. Note that although each
thread creates its own term, all terms are stored in the data structures in an intermixed way. 
Note that as there is no sharing here, each thread stores a full copy of the term in memory. 

The benchmark results in Table \ref{table:results_shared_lookup} and \ref{table:results_unique_lookup} were obtained by measuring the wall-clock time of creating $1000/\#\textit{threads}$ instances of the terms used in Table \ref{tabel:results_shared_create} and \ref{table:results_unique_create}. Before we start measuring the wall-clock times, the terms and subterms have already been inserted into the hash table, thus we are only measuring the cost of performing repeated lookups in our hash table.
The experiment reported in Table \ref{table:results_shared_lookup} is the same as the one in 
Table \ref{table:results_creating_shared_intel}, but the former is run on an AMD EPYC 7452 processor
whereas the latter uses an Intel i7-7700HQ processor.

The benchmark results in Table \ref{table:results_shared_inspect} were obtained by having each thread 
perform $1000/\#\textit{threads}$ breadth-first traversals of the term $t_{20}$ and measuring the wall-clock time. 
The traversal does not make use of the shared structure of terms, meaning that approximately $10^9$ term
nodes are visited. 
Similarly, the benchmark results in Table \ref{table:results_unique_aspect} were obtained by having each 
thread perform $1000/\#\textit{threads}$ breadth-first traversals of a term $t_{20}$ that is unique for each thread.

We also measured the wall-clock time of the state space generation of the 1394 firewire protocol using a parallel prototype of the mCRL2 toolset. The results are listed in Table \ref{table:result_statespace_exploration}.

\newcolumntype{C}{>{\centering}p{2.5em}}
\begin{table}[hb]
\resizebox{\textwidth}{!}{
\begin{tabular}{ |p{\widthof{sequential reference counter }} || *{10}{C|}c|}
\hline
 $\#\textit{Threads}$ &1 &2 &3 &4 &5 &6 &7 &8 &9 &10 &11 \\ \hline\hline
	parallel reference counter &
 61.0 & 87.6 & 90.1 & 83.1 & 67.0 & 57.4 & 56.2 & 53.3 & 50.0 & 46.5 & 43.4 \\
	parallel protection set &
 62.8 & 31.7 & 21.5 & 16.5 & 13.4 & 11.2 & 9.79 & 8.67 & 7.78 & 7.15 & 6.54 \\
	sequential reference counter &
 60.0 & & & & & & & & & & \\
	sequential protection set &
 52.9 & & & & & & & & & & \\
	original aterm library &
 40.2 & & & & & & & & & & \\
\hline\hline
 $\#\textit{Threads}$ &12 &13 &14 &15 &16 &17 &18 &19 &20 &21 &22 \\ \hline\hline
	parallel reference counter &
 43.3 & 41.4 & 38.8 & 38.1 & 37.3 & 35.5 & 35.2 & 35.2 & 33.9 & 33.2 & 32.5 \\
	parallel protection set &
 6.07 & 5.68 & 5.37 & 5.06 & 4.85 & 4.83 & 4.72 & 4.67 & 4.60 & 4.56 & 4.49 \\
\hline
\hline
 $\#\textit{Threads}$ &23 &24 &25 &26 &27 &28 &29 &30 &31 &32 &\\ \hline\hline
	parallel reference counter &
 31.7 & 31.3 & 30.8 & 29.4 & 28.7 & 28.1 & 28.0 & 27.9 & 27.0 & 26.6 &\\
	parallel protection set &
 4.44 & 4.37 & 4.32 & 4.23 & 4.17 & 4.15 & 4.11 & 4.07 & 4.02 & 3.99 &\\
\hline
\end{tabular}}
\caption{Wall-clock time for state space exploration.}
\label{table:result_statespace_exploration}
\end{table}

\begin{table}[h]
\centering
\begin{tabular}{| p{\widthof{sequential reference counter }} || *{7}{C|} c |}
\hline
 $\#\textit{Threads}$ &1 &2 &3 &4 &5 &6 &7 &8\\ \hline\hline
	parallel reference counter &
 10.0 & 10.4 & 4.42 & 4.08 & 3.22 & 2.74 & 2.38 & 2.22 \\
	parallel protection set &
 5.68 & 3.09 & 2.36 & 1.60 & 1.52 & 1.30 & 1.14 & 1.02 \\
	sequential reference counter &
 4.61 & & & & & & &  \\
	sequential protection set &
 4.20 & & & & & & & \\
	original aterm library &
 4.83 & & & & & & & \\
	parallel java &
 130 & 73.0 & 50.4 & 40.5 & 32.1 & 29.5 & 28.5 & 29.2 \\
	std::shared\_mutex &
 10.3 & 38.1 & 42.1 & 41.0 & 40.7 & 41.7 & 42.7 & 46.2 \\
 \hline
\end{tabular}
\caption{Wall-clock time for creating existing terms (shared, Intel).}
\label{table:results_creating_shared_intel}
\end{table}

\begin{table}[h]
\resizebox{\textwidth}{!}{
	\begin{tabular}{| p{\widthof{sequential reference counter }} || *{10}{C|}c|}
	\hline
 $\#\textit{Threads}$ &1 &2 &3 &4 &5 &6 &7 &8 &9 &10 &11 \\ \hline\hline
	parallel reference counter &
 0.03 & 0.11 & 0.22 & 0.14 & 0.26 & 0.34 & 0.28 & 0.29 & 0.39 & 0.51 & 0.53 \\
	parallel protection set &
 0.03 & 0.07 & 0.07 & 0.20 & 0.15 & 0.28 & 0.20 & 0.17 & 0.32 & 0.34 & 0.32 \\
	sequential reference counter &
 0.02 & & & & & & & & & & \\
	sequential protection set &
 0.02 & & & & & & & & & & \\
	original aterm library &
 0.01 & & & & & & & & & & \\
	parallel java &
 0.26 & 0.46 & 0.68 & 1.32 & 1.25 & 1.20 & 1.51 & 1.41 & 1.36 & 1.48 & 1.51 \\
	std::shared\_mutex &
 0.03 & 0.12 & 0.18 & 0.32 & 0.24 & 0.22 & 0.38 & 0.47 & 0.47 & 0.55 & 0.50 \\
 \hline\hline
 $\#\textit{Threads}$ &12 &13 &14 &15 &16 &17 &18 &19 &20 &21 &22 \\ \hline\hline
	parallel reference counter &
 0.51 & 0.59 & 0.50 & 0.54 & 0.49 & 0.53 & 0.54 & 0.48 & 0.5 & 0.48 & 0.52 \\
	parallel protection set &
 0.35 & 0.39 & 0.32 & 0.39 & 0.33 & 0.41 & 0.39 & 0.38 & 0.38 & 0.37 & 0.39 \\
	parallel java &
 1.40 & 1.51 & 1.44 & 1.43 & 1.38 & 1.67 & 1.77 & 1.87 & 1.77 & 1.78 & 1.85 \\
	std::shared\_mutex &
 0.58 & 0.62 & 0.63 & 0.67 & 0.72 & 0.74 & 0.76 & 0.79 & 0.83 & 0.83 & 0.88 \\
\hline\hline
 $\#\textit{Threads}$ &23 &24 &25 &26 &27 &28 &29 &30 &31 &32 &\\ \hline\hline
	parallel reference counter &
 0.50 & 0.53 & 0.50 & 0.49 & 0.53 & 0.51 & 0.51 & 0.49 & 0.48 & 0.48& \\
	parallel protection set &
 0.43 & 0.37 & 0.39 & 0.41 & 0.39 & 0.39 & 0.42 & 0.44 & 0.39 & 0.41& \\
	parallel java &
 1.68 & 1.95 & 1.68 & 1.82 & 1.81 & 1.65 & 1.94 & 1.94 & 2.09 & 1.86& \\
	std::shared\_mutex &
 0.91 & 0.95 & 0.96 & 0.99 & 0.98 & 1.04 & 1.02 & 1.10 & 1.11 & 1.16& \\
 \hline
\end{tabular}}
	\caption{Wall-clock time for creating new terms (shared).}
	\label{tabel:results_shared_create}
\end{table}

\begin{table}[h]
\resizebox{\textwidth}{!}{
\begin{tabular}{| p{\widthof{sequential reference counter }} || *{10}{C|}c|}
\hline
 $\#\textit{Threads}$ &1 &2 &3 &4 &5 &6 &7 &8 &9 &10 &11 \\ \hline\hline
	parallel reference counter &
 0.03 & 0.03 & 0.03 & 0.04 & 0.03 & 0.04 & 0.05 & 0.07 & 0.06 & 0.06 & 0.06 \\
	parallel protection set &
 0.03 & 0.05 & 0.03 & 0.04 & 0.03 & 0.03 & 0.04 & 0.05 & 0.05 & 0.05 & 0.06 \\
	sequential reference counter &
 0.02 & & & & & & & & & & \\
	sequential protection set &
0.02 & & & & & & & & & & \\
	original aterm library &
0.01 & & & & & & & & & & \\
	parallel java &
 0.26 & 0.25 & 0.28 & 0.26 & 0.28 & 0.28 & 0.28 & 0.35 & 0.33 & 0.35 & 0.32 \\
	std::shared\_mutex &
 0.03 & 0.04 & 0.04 & 0.05 & 0.10 & 0.04 & 0.04 & 0.09 & 0.09 & 0.09 & 0.09 \\
\hline \hline
 $\#\textit{Threads}$ &12 &13 &14 &15 &16 &17 &18 &19 &20 &21 &22 \\ 
 \hline\hline
	parallel reference counter &
 0.06 & 0.05 & 0.05 & 0.07 & 0.06 & 0.06 & 0.05 & 0.06 & 0.07 & 0.06 & 0.06 \\
	parallel protection set &
 0.05 & 0.06 & 0.06 & 0.05 & 0.06 & 0.07 & 0.06 & 0.05 & 0.06 & 0.06 & 0.06 \\
	parallel java &
 0.33 & 0.35 & 0.32 & 0.34 & 0.33 & 0.34 & 0.33 & 0.35 & 0.33 & 0.34 & 0.34 \\
	std::shared\_mutex &
 0.09 & 0.10 & 0.10 & 0.10 & 0.10 & 0.09 & 0.11 & 0.09 & 0.12 & 0.11 & 0.13 \\
\hline
\hline
 $\#\textit{Threads}$ &23 &24 &25 &26 &27 &28 &29 &30 &31 &32 &\\ 
 \hline\hline
	parallel reference counter &
 0.07 & 0.06 & 0.05 & 0.06 & 0.06 & 0.07 & 0.07 & 0.07 & 0.06 & 0.06 &\\
	parallel protection set &
 0.06 & 0.06 & 0.05 & 0.07 & 0.06 & 0.06 & 0.05 & 0.05 & 0.05 & 0.06 &\\
	parallel java &
 0.37 & 0.34 & 0.35 & 0.35 & 0.37 & 0.35 & 0.34 & 0.34 & 0.35 & 0.34 &\\
	std::shared\_mutex &
 0.13 & 0.11 & 0.14 & 0.09 & 0.15 & 0.18 & 0.15 & 0.13 & 0.16 & 0.13 &\\
 \hline
\end{tabular}}

\caption{Wall-clock time for creating new terms (distinct).}
\label{table:results_unique_create}
\end{table}

\begin{table}[h]
\resizebox{\textwidth}{!}{
\begin{tabular}{| p{\widthof{sequential reference counter }} || *{10}{C|}c|}
\hline
 $\#\textit{Threads}$ &1 &2 &3 &4 &5 &6 &7 &8 &9 &10 &11 \\ \hline\hline
	parallel reference counter &
 9.01 & 35.1 & 11.5 & 19.0 & 9.81 & 10.6 & 6.40 & 6.00 & 4.90 & 4.85 & 3.88 \\
	parallel protection set &
 4.07 & 2.51 & 1.66 & 1.39 & 1.07 & 0.96 & 0.81 & 0.75 & 0.67 & 0.59 & 0.53 \\
	sequential reference counter &
 4.01 & & & & & & & & & & \\
	sequential protection set &
 3.65 & & & & & & & & & & \\
	original aterm library &
 4.57 & & & & & & & & & & \\
	parallel java &
 104 & 136 & 106 & 103 & 91.7 & 84.2 & 75.5 & 71.1 & 64.5 & 61.6 & 57.5 \\
	std::shared\_mutex &
 6.51 & 14.2 & 15.1 & 15.1 & 18.7 & 20.7 & 22.0 & 33.3 & 28.1 & 25.4 & 24.5 \\
\hline
\hline
 $\#\textit{Threads}$ &12 &13 &14 &15 &16 &17 &18 &19 &20 &21 &22 \\ 
 \hline\hline
	parallel reference counter &
 3.51 & 3.10 & 2.86 & 2.77 & 2.66 & 2.45 & 2.61 & 2.43 & 2.33 & 2.27 & 2.17 \\
	parallel protection set &
 0.49 & 0.46 & 0.43 & 0.41 & 0.40 & 0.39 & 0.37 & 0.36 & 0.35 & 0.32 & 0.31 \\
	parallel java &
 54.3 & 51.9 & 49.4 & 48.7 & 46.5 & 47.7 & 47.9 & 48.1 & 48.6 & 47.7 & 46.6 \\
	std::shared\_mutex &
 22.8 & 22.7 & 23.1 & 22.6 & 22.5 & 22.9 & 23.2 & 23.7 & 24.7 & 24.1 & 24.3 \\
\hline
\hline
 $\#\textit{Threads}$ &23 &24 &25 &26 &27 &28 &29 &30 &31 &32 &\\ 
 \hline\hline
	parallel reference counter &
 2.16 & 2.10 & 2.08 & 2.04 & 2.03 & 1.97 & 1.82 & 1.87 & 1.81 & 1.81 &\\
	parallel protection set &
 0.32 & 0.31 & 0.29 & 0.3 & 0.28 & 0.27 & 0.29 & 0.28 & 0.28 & 0.29 &\\
	parallel java &
 45.9 & 45.4 & 45.8 & 44.9 & 44.8 & 42.4 & 42.4 & 42.9 & 42.1 & 42.1 &\\
	std::shared\_mutex &
 24.7 & 24.4 & 25.0 & 25.1 & 25.4 & 25.5 & 26.0 & 26.7 & 27.5 & 28.3 &\\
 \hline
\end{tabular}}
\caption{Wall-clock time for creating existing terms (shared).}
\label{table:results_shared_lookup}
\end{table}
	
\begin{table}[h]
\resizebox{\textwidth}{!}{
\begin{tabular}{| p{\widthof{sequential reference counter }} || *{10}{C|}c|}
\hline
 $\#\textit{Threads}$ &1 &2 &3 &4 &5 &6 &7 &8 &9 &10 &11 \\ \hline\hline
	parallel reference counter &
 9.02 & 5.15 & 3.95 & 3.05 & 2.64 & 2.25 & 2.38 & 2.41 & 2.42 & 2.49 & 2.24 \\
	parallel protection set &
 3.96 & 2.66 & 2.58 & 2.44 & 2.27 & 1.76 & 1.92 & 1.86 & 1.95 & 1.91 & 1.79 \\
	sequential reference counter &
 4.02 & & & & & & & & & & \\
	sequential protection set &
 3.71 & & & & & & & & & & \\
	original aterm library &
 4.58 & & & & & & & & & & \\
 	parallel java &
 106 & 212 & 218 & 227& 260 & 266 &  274 & 276 & 295 & 272 & 287 \\ 
	std::shared\_mutex &
 6.46 & 13.8 & 15.5 & 15.7 & 18.3 & 18.5 & 24.6 & 33.2 & 26.9 & 25.0 & 23.7 \\
\hline\hline
 $\#\textit{Threads}$ &12 &13 &14 &15 &16 &17 &18 &19 &20 &21 &22 \\ 
 \hline\hline
	parallel reference counter &
 2.22 & 2.27 & 2.26 & 2.20 & 2.39 & 2.55 & 2.63 & 2.58 & 2.72 & 2.67 & 2.67 \\
	parallel protection set &
 1.78 & 1.76 & 1.77 & 1.75 & 1.81 & 1.82 & 1.97 & 2.05 & 2.04 & 2.07 & 2.07 \\
	parallel java& 
 275 & 287 & 296 & 2912 & 281 & 286 & 280 & 284 & 294 & 292 & 314\\
	std::shared\_mutex &
 22.7 & 22.4 & 22.9 & 22.6 & 22.2 & 22.7 & 23.2 & 23.8 & 24.7 & 24.1 & 24.1 \\
\hline\hline
 $\#\textit{Threads}$ &23 &24 &25 &26 &27 &28 &29 &30 &31 &32 &\\ 
 \hline\hline
	parallel reference counter &
 2.69 & 2.69 & 2.77 & 2.76 & 2.8 & 2.77 & 2.82 & 2.84 & 2.86 & 2.92 &\\
	parallel protection set &
 2.07 & 2.10 & 2.15 & 2.05 & 2.12 & 2.11 & 2.17 & 2.22 & 2.27 & 2.22 &\\
 	parallel java &
 	311 & 308 & 315 & 316 & 324 & 330 & 339 & 332 & 343 & 352&\\
	std::shared\_mutex &
 24.4 & 24.4 & 25.3 & 25.3 & 25.9 & 25.7 & 26.3 & 27.1 & 27.8 & 28.7 &\\
\hline
\end{tabular}}
\caption{Wall-clock time for creating existing terms (distinct).}
\label{table:results_unique_lookup}
\end{table}

\begin{table}[h]
\resizebox{\textwidth}{!}{
\begin{tabular}{| p{\widthof{sequential reference counter }} || *{10}{C|}c|}
\hline
 $\#\textit{Threads}$ &1 &2 &3 &4 &5 &6 &7 &8 &9 &10 &11 \\ \hline\hline
	parallel reference counter &
 15.7 & 8.63 & 5.93 & 4.60 & 3.87 & 3.41 & 3.00 & 2.79 & 2.50 & 2.45 & 2.21 \\
	parallel protection set &
 16.7 & 8.90 & 6.07 & 4.66 & 3.93 & 3.37 & 3.01 & 2.80 & 2.55 & 2.41 & 2.34 \\
	sequential reference counter &
 16.8 & & & & & & & & & & \\
	sequential protection set &
 18.2 & & & & & & & & & & \\
	original aterm library &
 16.4 & & & & & & & & & & \\
	parallel java &
 34.6 & 34.5 & 36.0 & 36.7 & 36.1 & 33.6 & 30.9 & 28.4 & 26.4 & 25.0 & 22.9 \\
	std::shared\_mutex &
 16.2 & 8.71 & 5.95 & 4.54 & 3.86 & 3.34 & 3.01 & 2.74 & 2.53 & 2.40 & 2.29 \\
\hline\hline
 $\#\textit{Threads}$ &12 &13 &14 &15 &16 &17 &18 &19 &20 &21 &22 \\ \hline\hline
	parallel reference counter &
 2.17 & 2.21 & 2.23 & 2.32 & 2.24 & 2.14 & 2.30 & 2.21 & 2.11 & 2.21 & 2.09 \\
	parallel protection set &
 2.28 & 2.21 & 2.30 & 2.35 & 2.21 & 2.26 & 2.35 & 2.25 & 2.33 & 2.28 & 2.21 \\
	parallel java &
 17.8 & 20.6 & 17.7 & 19.1 & 22.3 & 19.5 & 19.6 & 20.4 & 21.9 & 21.6 & 21.5 \\
	std::shared\_mutex &
 2.25 & 2.33 & 2.28 & 2.24 & 2.21 & 2.22 & 2.29 & 2.15 & 2.24 & 2.04 & 2.24 \\
\hline\hline
 $\#\textit{Threads}$ &23 &24 &25 &26 &27 &28 &29 &30 &31 &32 &\\ \hline\hline
	parallel reference counter &
 2.17 & 2.18 & 2.13 & 2.07 & 2.06 & 2.09 & 2.06 & 2.05 & 2.13 & 2.10 &\\
	parallel protection set &
 2.26 & 2.15 & 2.12 & 2.16 & 2.13 & 2.07 & 2.09 & 2.16 & 2.03 & 2.03 &\\
	parallel java &
 17.6 & 19.2 & 18.4 & 20.6 & 22.7 & 20.8 & 18.5 & 22.5 & 23.6 & 19.4 &\\
	std::shared\_mutex &
 2.24 & 2.12 & 2.18 & 2.05 & 2.05 & 2.2 & 2.16 & 2.17 & 2.02 & 2.07 &\\
 \hline
\end{tabular}}
\caption{Wall-clock time for traversing terms (shared).}
\label{table:results_shared_inspect}
\end{table}

\begin{table}[h]
\resizebox{\textwidth}{!}{
\begin{tabular}{| p{\widthof{sequential reference counter }} || *{10}{C|}c|}
\hline
 $\#\textit{Threads}$ &1 &2 &3 &4 &5 &6 &7 &8 &9 &10 &11 \\ \hline\hline
	parallel reference counter &
 18.4 & 9.61 & 6.41 & 4.93 & 4.10 & 3.56 & 3.14 & 2.88 & 2.63 & 2.51 & 2.34 \\
	parallel protection set &
 17.0 & 8.80 & 6.03 & 4.59 & 3.85 & 3.39 & 3.00 & 2.78 & 2.56 & 2.40 & 2.28 \\
	sequential reference counter &
 15.9 & & & & & & & & & & \\
	sequential protection set &
 18.3 & & & & & & & & & & \\
	original aterm library &
 17.4 & & & & & & & & & & \\
	parallel java &
 34.5 & 34.2 & 35.8 & 37.0 & 35.5 & 33.8 & 30.1 & 28.3 & 27.1 & 23.4 & 21.9 \\
	std::shared\_mutex &
 16.5 & 8.59 & 5.98 & 4.63 & 3.99 & 3.47 & 3.07 & 2.88 & 2.60 & 2.49 & 2.43 \\
\hline\hline
 $\#\textit{Threads}$ &12 &13 &14 &15 &16 &17 &18 &19 &20 &21 &22 \\ \hline\hline
	parallel reference counter &
 2.36 & 2.33 & 2.33 & 2.28 & 2.24 & 2.19 & 2.26 & 2.22 & 2.16 & 2.20 & 2.11 \\
	parallel protection set &
 2.31 & 2.38 & 2.28 & 2.31 & 2.20 & 2.21 & 2.21 & 2.13 & 2.21 & 2.16 & 2.18 \\
	parallel java &
 21.1 & 17.5 & 20.9 & 17.3 & 18.7 & 20.8 & 23.0 & 18.4 & 21.6 & 23.0 & 22.8 \\
	std::shared\_mutex &
 2.56 & 2.52 & 2.50 & 2.41 & 2.32 & 2.25 & 2.4 & 2.37 & 2.25 & 2.34 & 2.39 \\
\hline\hline
 $\#\textit{Threads}$ &23 &24 &25 &26 &27 &28 &29 &30 &31 &32 &\\ \hline\hline
	parallel reference counter &
 2.27 & 2.16 & 2.16 & 2.13 & 2.10 & 2.34 & 2.13 & 2.06 & 2.16 & 2.05 &\\
	parallel protection set &
 2.27 & 2.15 & 2.16 & 2.17 & 2.12 & 2.08 & 2.11 & 2.10 & 2.06 & 2.04 &\\
	parallel java &
 18.3 & 22.2 & 22.3 & 22.5 & 18.7 & 21.7 & 22.2 & 19.3 & 22.8 & 19.5 &\\
	std::shared\_mutex &
 2.18 & 2.30 & 2.27 & 2.34 & 2.14 & 2.28 & 2.08 & 2.10 & 2.38 & 2.26 &\\
 \hline
\end{tabular}}
\caption{Wall-clock time for traversing terms (distinct).}
\label{table:results_unique_aspect}
\end{table}

\end{document}